\pdfoutput=1
\documentclass[usenatbib,fleqn]{mnras}
 \usepackage{ifthen} \newboolean{pubformat} \setboolean{pubformat}{true}
\usepackage{natbib}		
\usepackage[draft]{pdfpages}
\newcommand{\continuefloat}
{\ifthenelse{\boolean{pubformat}}{\contcaption{}}{\ContinuedFloat}}
\ifthenelse{\boolean{pubformat}}
{
\usepackage{fixltx2e}     
\newcommand{\onecol}[1]{\includegraphics[width=84mm]{#1}}

}{
\usepackage{times}
\newcommand{\sun}{\odot}
\newcommand{\onecol}[1]{\includegraphics[width=6.5in]{#1}}

}
\usepackage{graphicx}
\title[Orbital timescales in the Sagittarius Stream]
{Connecting the Milky Way potential profile to the orbital timescales
and spatial structure of the Sagittarius Stream}
\author[M. A.\ Fardal et al.]   
{Mark A.\ Fardal$^{1,2}$\thanks{E-mail: fardal@stsci.edu}, 
Roeland P. van der Marel$^{1,3}$,
David R. Law$^1$,\newauthor
Sangmo Tony Sohn$^1$,
Branimir Sesar$^4$,
Nina Hernitschek$^5$,
Hans-Walter Rix$^4$\\
$^1$Space Telescope Science Institute, 3700 San Martin Drive, Baltimore, MD 21218, USA\\
$^2$Department of Astronomy, University of Massachusetts, Amherst, MA 01003-9305, USA\\
$^3$Center for Astrophysical Sciences, Department of Physics \& Astronomy,
Johns Hopkins University, Baltimore, MD 21218, USA\\
$^4$Max Planck Institute for Astronomy, K\"{o}enigstuhl 17, D-69117, Heidelberg, Germany\\
$^5$Division of Physics, Mathematics and Astronomy, Caltech, Pasadena, CA 91125
}  
\date{Draft v0.9.X, \today}
\pagerange{\pageref{firstpage}--\pageref{lastpage}} \pubyear{2018} 
\ifthenelse{\boolean{pubformat}}
{
\voffset -1.4cm    
\setlength{\textheight}{241mm}  
}{
\setlength{\textwidth}{6.5in}
\setlength{\oddsidemargin}{0in}
\setlength{\evensidemargin}{0in}
\setlength{\textheight}{227.6mm}  
\topmargin=0mm
\headheight=0mm
\headsep=0mm
\footskip=13mm
}

\newcommand{\Myr}{\,\mbox{Myr}}

\newcommand{\kpc}{\,\mbox{kpc}}

\newcommand{\kms}{\,\mbox{km}\,\mbox{s}^{-1}}

\newcommand{\msun}{\,M_{\sun}}
\newcommand{\degree}{\degr}
\newcommand{\nbody}{$N$-body}
\newcommand{\tsim}{\sim\!}

\newcommand{\ltrsim}{\la}
\defcitealias{sesar17}{S17}
\defcitealias{law10}{LM10}
\defcitealias{dierickx17}{D17}
\begin{document}
\maketitle    
\label{firstpage}
\begin{abstract}
Recent maps of the halo using RR Lyrae from Pan-STARRS1 depict the
spatial structure of the Sagittarius stream, showing the leading and
trailing stream apocenters differ in galactocentric radius by a factor
of two, and also resolving substructure in the stream at these
apocenters.  Here we present dynamical models that reproduce these
features of the stream in simple Galactic potentials.  We find that
debris at the apocenters must be dynamically young, being stripped off 
in pericentric passages either one or two orbital periods ago.  The ratio of
leading and trailing apocenters is sensitive to both dynamical friction and the 
outer slope of the Galactic rotation curve.  These dependencies can be understood
with simple regularities connecting the apocentric radii, circular
velocities, and orbital period of the progenitor.  The effect of
dynamical friction can be constrained using substructure within the
leading apocenter.  Our ensembles of models are not statistically
proper fits to the stream.  Nevertheless, we consistently find the
mass within 100 kpc to be $\sim \! 7 \times 10^{11} \, M_{\odot}$, with a
nearly flat rotation curve between 50 and 100 kpc.  This points to a
more extended Galactic halo than assumed in some current models.  We
show one example model in various observational dimensions.  A plot of
velocity versus distance separates younger from older debris,
and suggests that the young trailing debris will serve as an
especially useful probe of the outer Galactic potential.
\end{abstract}
\section{INTRODUCTION}
If one attempted to design a tidal stream to probe the potential of the outer Milky Way
potential, it would probably look much like the Sagittarius (Sgr)
dwarf galaxy's stellar stream.  This majestic 
structure contains nearly $10^9 \msun$ in stars with many useful tracers among them,
shows at least three and perhaps more radial turning points,
and wraps more than one full circle around the Galaxy.   As the number of large-area surveys
continues to grow, we are gaining an increasingly clear view of the spatial extent
and kinematics of this object. (See \citealp{law16} for a comprehensive review.)
It would seem to follow that we are thereby gaining an increasingly clear view of the
outer Milky Way halo potential.

Unfortunately, our understanding of the stream's dynamics has not kept
up with the observations.  By now there are several long-standing problems
with Sagittarius stream models, most famously in the leading stream where velocities
favor a prolate halo and the stream latitude favors an oblate halo
\citep{helmi04,johnston05,law05}.
Also, a ``bifurcation'' of the stream in latitude is seen now in both the leading and
trailing arms \citep{belokurov06,koposov12}.
Recently, another important issue has been
created by observations near the trailing apocenter.
This shows the turnaround of the trailing stream occurs at 
galactocentric radius $R_{GC} = 100 \kpc$
\citep{newberg03,drake13,belokurov14,sesar17},
Furthermore, the high-contrast three-dimensional
view given by the RR Lyraes in Pan-STARRS1 
(\citealp[][henceforth S17]{sesar17}, and \citealp{hernitschek17})
suggests that both the leading and trailing apocenters have two
components at slightly different distances (their fig.~1).

The stream properties at apocenter are closely linked to its orbital energy,
and the substructure is presumably created by two different pericentric passages.
Thus these observables are perhaps the most basic aspects to get right when
modeling the stream.
Perhaps the best-known model is that of \citetalias{law10};
but the trailing apocenter in this model reaches only to a galactocentric radius of about 65 kpc.
\citet{gibbons14} discuss modeling of the stream explicitly aimed at reproducing
the apocentric properties, and infer the Galactic potential to have a lower
mass than in the \citetalias{law10} model.  But while this paper includes parameter inference, it
does not include model plots or statistical tests that would
make clear whether their model is a good fit to the data.
Finally, the model of \citet[][henceforth D17]{dierickx17} perhaps comes
the closest in reproducing the apocentric
properties---rather impressively because it was not an actual fit to the stream---but
many aspects of the stream such as its sky orientation, apocenter radii, and so on
are only loosely matched.

In this paper we aim to take another step towards understanding the Sagittarius
Stream and the Milky Way halo, by modeling the spatial and kinematic properties of the
stream with a focus on the apocentric features seen in the \citetalias{sesar17} data.
We will not address every feature of the stream with the same attention.
Nor do we draw rigorous statistical inferences about parameters of our model.
As past literature on this stream may suggest, we believe that
such inferences are often highly model-dependent and still premature.
Rather, we focus here on drawing out features of the models that are helpful in
fitting the spatial properties of the leading and trailing arms.

The paper is laid out as follows.
In Section~\ref{sec.obs_structure}, we provide a quick overview of the observed structure
of the stream, illustrating the specific features used to constrain the
models.
In Section~\ref{sec.methods},
we explain the dynamical modeling techniques,
satellite structure, Milky Way potentials, and observational data
we use to generate models of the stream.
In Section~\ref{sec.modelphysics}, to gain understanding,
we first discuss the behavior of models in Milky Way potentials of differing
radial profiles that are all either spherical or very nearly so.
We will see that the ratio of apocenters is highly sensitive to the
outer slope of the rotation curve, 
through its effect on the orbital period of the stars
in the leading and trailing streams.
We next examine the influence of dynamical friction on the stream models.
Finally, we relax the spherical constraint and illustrate the resulting changes in the models.
Section~\ref{sec.modelprops} depicts the model properties in more detail,
comparing to observations and suggesting ways to detect and
make use of substructure in the stream.
Section~\ref{sec.discussion} discusses the relationship of our results to previous models, 
and addresses issues to be confronted in future observational and theoretical work. 
Finally, Section~\ref{sec.conclusions} summarizes our conclusions.
\section{Observed Structure and Substructure of the Stream}
\label{sec.obs_structure}
\begin{figure}
\onecol{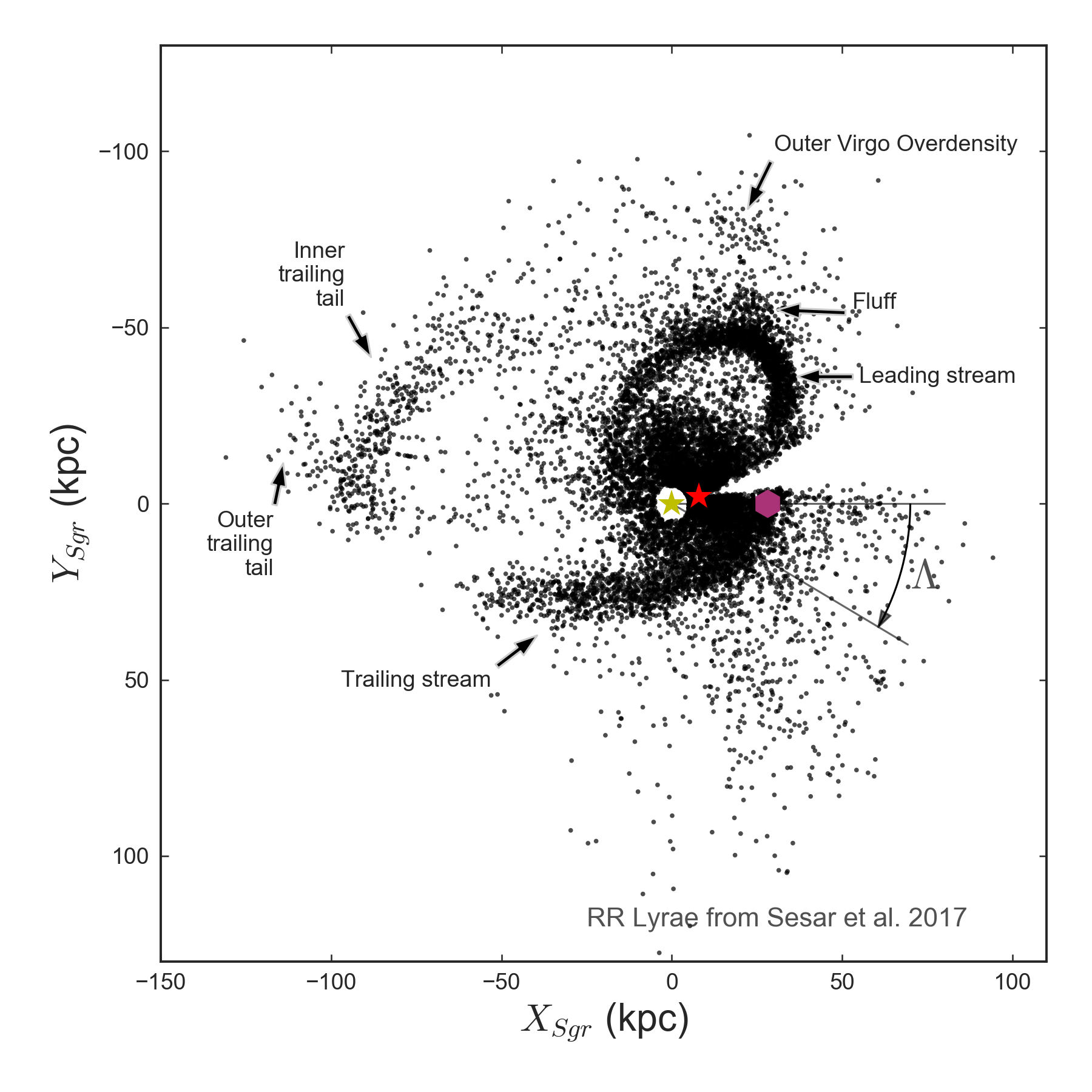}
\caption{
\label{fig.obs_structure}
Observed structure of the Sagittarius stream plotted in its orbital plane, and centered
on the Sun.
Points show RR Lyraes from \citetalias{sesar17} within $13\deg$ of the Sgr plane.
The missing wedge slightly inclined to the $X_{Sgr}$ axis is heavily 
incomplete due to Galactic extinction.
The yellow star indicates the Sun, and the red diamond the Galactic center.
The Sagittarius dSph (purple hexagon) lies along the $X_{Sgr}=0$ line at a
somewhat uncertain distance of $\tsim 28 \kpc$.
The arrow shows the origin and direction of stream longitude $\Lambda$,
defined so it increases opposite the actual motion of the stream.
We have noted several structures discussed in the text.
}
\end{figure}

Fig.~\ref{fig.obs_structure} reproduces the spatial positions of RR Lyraes in the
Sagittarius orbital plane from
the catalog of \citet{sesar17cat},
which has recently also been analyzed in more
detail by \citetalias{sesar17} and \citet{hernitschek17}.
We also mark the position of the Sun (at the origin in this plot),
the Galactic center, and the Sagittarius dSph galaxy itself.
The plot uses stars 
within $13\degr$ of the Sgr orbital plane
and with RR Lyrae classification score $\mathtt{score}_{3,ab} > 0.8$.
The catalog has now been published as table 1 of the 
electronic version of \citet{sesar17cat}.
Observations suggest that the dSph is currently close to pericenter.
In later analysis, we will use the Sagittarius coordinate system
defined by \citet{majewski03} and \citetalias{law10}.
Here longitude $\Lambda$ is zero at the position of the dSph and
increases in the direction {\em opposite} to the motion of the stream.
Latitude $B$ is defined using a left-handed system.
Then $X_{Sgr}$ is roughly (to within $15 \degree$)
aligned with galactic $X$ and $Y_{Sgr}$ with galactic $-Z$.
We have marked the leading and trailing Sgr streams
and several indications of substructure within the stream.
Aside from the indicated features, much of the structure visible in
the plot is probably unrelated to the Sgr galaxy,
although our understanding of halo substructure is evolving rapidly.
One stream feature we
will try to reproduce in this paper is the large difference between the radii
of the leading ($\tsim 50 \kpc$) and trailing ($\tsim 100 \kpc$) apocenters.
Indications of such a large difference have
built up gradually for more than a decade, but only
recently have become completely clear.

\begin{figure}
\onecol{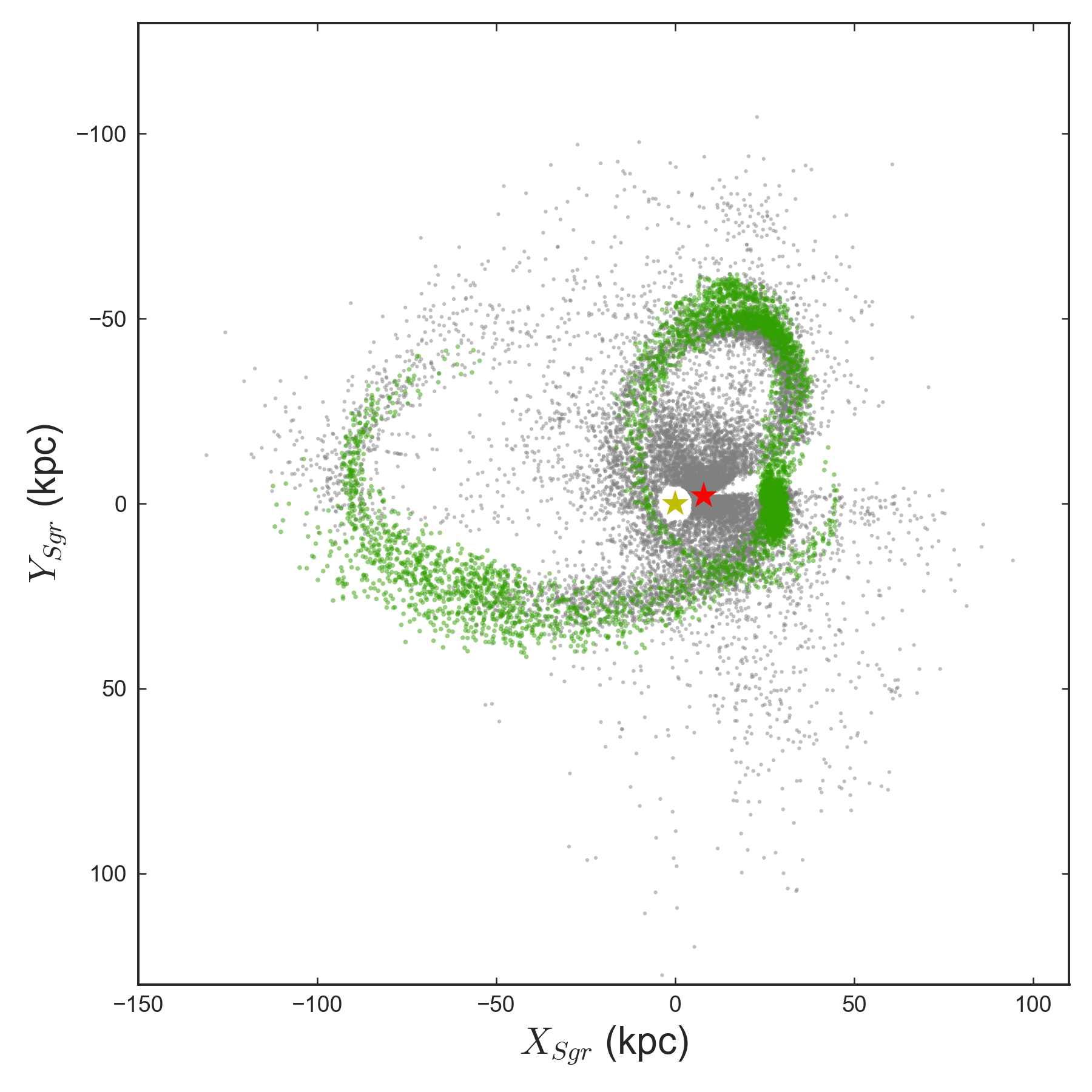}
\caption{
\label{fig.nbodyex}
Structure of stream \nbody\ model A in the Sagittarius plane,
shown with green points.
Gray points show the RR Lyraes from Fig.~\ref{fig.obs_structure}.
}
\end{figure}
\begin{figure}
\onecol{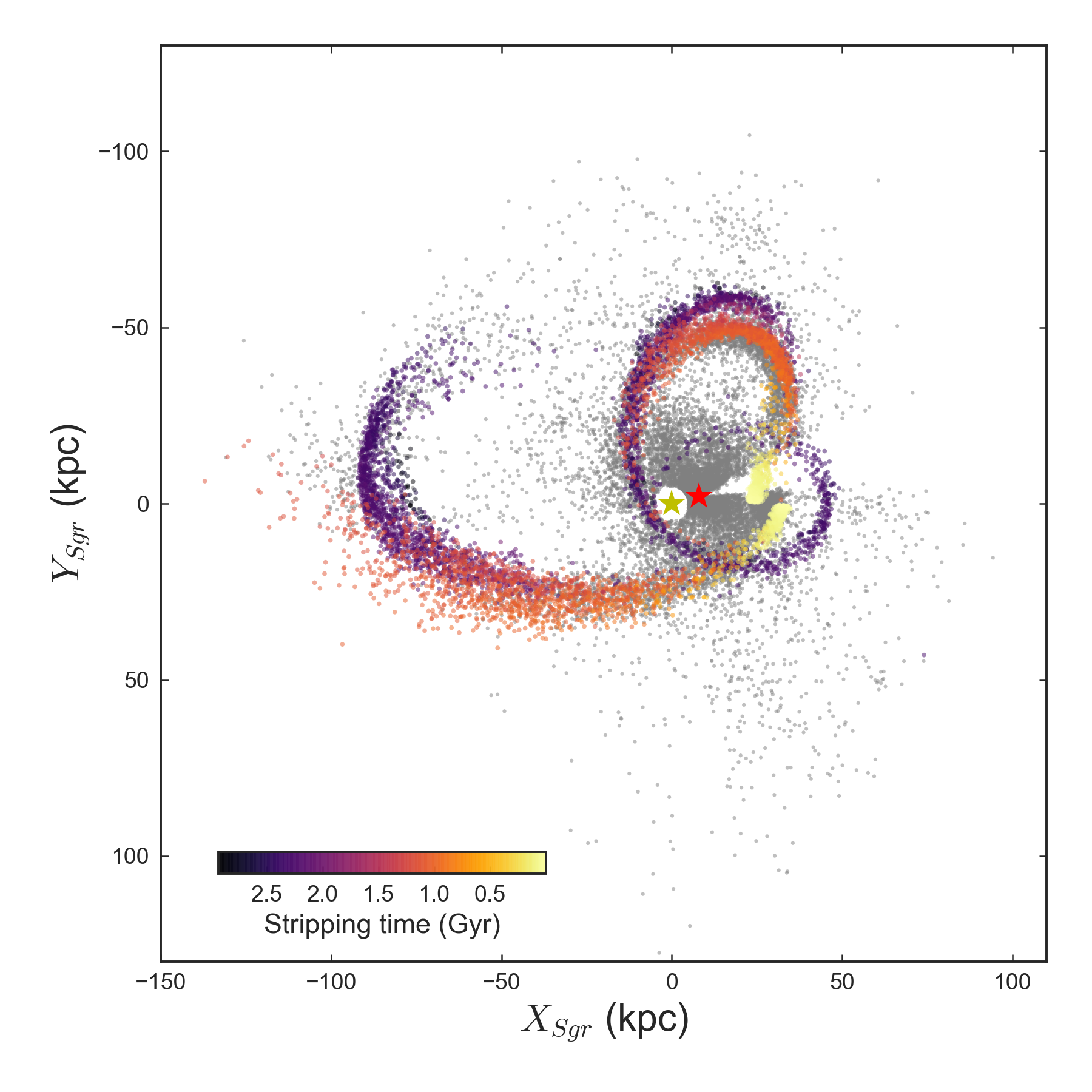}
\caption{
\label{fig.sprayex}
Structure of the particle-spray model A in the Sagittarius plane.
Model parameters are the same as in Fig.~\ref{fig.nbodyex}.
Point color corresponds to the lookback time of ejection from the satellite,
with lighter colors ejected more recently.
Gray points again show the RR Lyraes from Fig.~\ref{fig.obs_structure}.
}
\end{figure}

In this paper, we will use the term ``apocenter'' somewhat loosely to
mean the point where the observed distance to the stream turns around,
either in heliocentric or galactic coordinates.  It is clear that the
choice of origin will somewhat affect both the azimuth and distance 
to the turnaround point.  Furthermore, even the galactocentric
turnaround does not coincide with the actual apocenter of stars in the
stream.  In fact the stars at the stream turnaround are generally
outflowing, which produces an outward drift of the turnaround with time.
However, all of these turnaround points are close to each other and the
exact meaning should be clear in context.

To clarify our interpretation of the observed structure,
Figs.~\ref{fig.nbodyex} and \ref{fig.sprayex}
give examples of model fits to the stream.
These models have already been tuned to match the stream in the manner described later.
In Fig.~\ref{fig.sprayex} we color-code the models by the age of the debris,
defined here (and throughout the paper)
by the time since the stars were stripped from the satellite,
not by the ages of the stars themselves.
One might though reasonably expect 
some correlation between those two timescales, in view of Sgr's very extended
star formation history \citep{law16}.

A second important feature we aim to reproduce is the substructure
within the stream shown by the 
\citetalias{sesar17} data projected onto the Sgr orbital plane.
The trailing stream shows a clear dichotomy between an inner
tail that curves gently around apocenter
(``Feature 1'' in \citetalias{sesar17}), and an outer tail
that extends straight outwards with no sign of return
(``Feature 2'' in \citetalias{sesar17}).  These two components appear
very similar to stream models where the trailing debris comes from the
last two pericentric passages, as in Figs.~\ref{fig.nbodyex} and \ref{fig.sprayex}.
The straight outer stream in these models is the younger component, and the curved one
the older.  
Indeed, the resemblance is so close that we consider no alternative explanations in this paper.
(As discussed below and in \citetalias{sesar17},
the same feature is also seen in some earlier models of the Sagittarius stream
including that of \citetalias{dierickx17}.)

At the leading apocenter, there is some ``fluff'' visible
lying outside the main component of the stream (``Feature 3'' in \citetalias{sesar17}).
The presence of this feature has recently been verified using blue horizontal branch stars
by \citet{fukushima17}.
Given the close agreement in radius and latitude, and the otherwise low density of
halo RR Lyraes at this distance, it seems highly likely that this fluff is associated
with the Sagittarius stream.
It plausibly represents an older component of stream
debris at larger orbital energy, as illustrated in Figs.~\ref{fig.nbodyex} and \ref{fig.sprayex}.
There are some differences in morphology that might disfavor this explanation.
The fluff is concentrated in one spatial region and appears
more spatially concentrated than in the models.
This morphology is better matched in the \nbody\ run, though there is a
slightly offset in position in the similar feature there.
At present it seems reasonable that differences could be explained either
by slight changes to the basic dynamical model,
statistical fluctuations in the stars, or
varying spatial completeness of the \citetalias{sesar17} data.
Thus at present we regard an older, higher-energy stream component as
the most natural explanation of this structure.  

There is another intriguing structure in the vicinity of the leading apocenter
as indicated in Fig.~\ref{fig.obs_structure}, termed the ``outer Virgo overdensity''
or ``Feature 4'' in \citetalias{sesar17}.  
This lies about $9\degree$ off the Sagittarius orbital plane,
at a similar azimuth but a higher galactocentric radius
of about 80~kpc (and a similar heliocentric distance).
Later we will consider what would be necessary to produce such a component
within a Sagittarius stream model.  For now, we regard a direct association as less likely
than for the ``fluff'' component.

If either of these observed structures represents an older, higher-energy stream component,
then the main body of the stream at leading apocenter {\em must} be from younger, lower-energy
debris, presumably stripped only one or two Sagittarius dSph orbital periods ago.
This would make the dominant leading apocenter component younger than the
old debris that appears to dominate at trailing apocenter.
In the models, the difference in which stream is dominant
is caused by the shorter orbital times of the leading apocenter debris.

We obtain distances to the stream from the RR Lyrae sample of \citetalias{sesar17}.
Recently quantitative fits to the stream have also been presented by
\citet{hernitschek17}.  
However, by the time this paper appeared we had already conducted much
of the modeling work with our own fits to the same catalogue.
The Hernitschek fits extend over a wider longitude range and include quantities
such as the distance dispersion and amplitude.  
Our fits though have the advantage that they specifically fit to the main body of
the stream in cases where more than one component is visible.
The results are still very close to those of \citet{hernitschek17}.
Hence we persist with our own fitted values as the starting point for the modeling.
We use the catalogue of stars plotted in Fig.~1 of \citetalias{sesar17}.
We use only stars in three longitude segments of interest, namely those 
near leading and trailing apocenter and along the southern trailing stream.
We bin the stars by longitude $\Lambda$, and after plotting the stars in each bin
judge whether one or two components are detectable.
Only at leading and trailing apocenter do we see two components,
and in both cases the component at smaller distance is the stronger peak
that we use as the stream distance.
We then perform a maximum likelihood fit to the data in each bin,
using a distribution that assumes one or two components as appropriate
plus a linear baseline component.
We list distance estimates and formal errors from the fit in Table~\ref{tab.distance}.
They are shown graphically in Section 5 below (see Figure 14).

Using these fits, the maximum galactocentric distance of the stream appears to be 49 and
101 kpc at leading and trailing apocenter respectively.
These numbers are close to the values of 
$47.8 \pm 0.5 \kpc$ and $102.5 \pm 2.5 \kpc$ obtained by \citealp{belokurov14}
using horizontal branch stars.
The uncertainties on our values
are large enough that we will use the round numbers 50 and 100 kpc for simplicity.
The apocenters thus differ by a factor of two.  
\section{Modeling ingredients}
\label{sec.methods}
\subsection{Basic assumptions}
\label{sec.assumptions}
We use two types of dynamical models in this paper: standard \nbody\ models,
and particle spray models.  The latter type was pioneered by
\citet{kuepper08}, with further development under different names in numerous papers
\citep{varghese11,kuepper12,gibbons14,bonaca14}.
The specific recipe used here is presented and tested against \nbody\ models
in \citet{fardal15}.
Figs.~\ref{fig.nbodyex} and \ref{fig.sprayex}
show examples of each type, together with the RR Lyrae data from \citetalias{sesar17}.
These two models are run with the same center-of-mass orbital parameters
of the progenitor and Milky Way potential,
and are thus directly comparable.  Both models demonstrate reasonable agreement with
the spatial structure of the stream.
In the spray run we explicitly assign the times
at which particles were ejected from the satellite
(as shown in Fig.~\ref{fig.sprayex})
and compute various other properties of the orbit in the host potential.  This allows
more detailed examination of the debris properties than in the \nbody\ runs,
where we only retain the final phase space and do not track the ejection time
or other orbital properties.

As already mentioned, the primary observational feature we aim to model
is the large ratio of trailing to leading apocentric radii.
Generically, the leading stream is more tightly bound to the host
(i.e., has a lower orbital energy) than the trailing stream.
Assuming a potential that is close to spherical at these large radii,
it therefore has a smaller apocentric
radius than the trailing stream
(see \citealp{law05}; \citetalias{law10}).
However, reproducing the factor of two difference
seen in the case of Sagittarius is a challenge not met by current models.

The differences between the stream and the orbit are increased,
and thus so is the ratio of apocenter radii,
when the debris is young, in the sense of having experienced few orbital periods.
We interpret the split in the trailing stream near apocenter as a clear sign
that the Sagittarius dSph has experienced {\em at least} two pericentric passages that resulted
in major disruption and formed its stream.
Here we are not counting the current, incomplete pericentric passage,
which is expected to contribute only short tidal features incapable
of reaching all the way to apocenter at present. (See for example the lightest yellow
points in Fig.~\ref{fig.sprayex}.)
We can thus maximize the apocenter ratio by assuming that there have been in fact
{\em only} two disruptive pericenters (again excluding the current one).

We have tested models where the debris near the apocenters results from 
earlier pericentric passages, and have
consistently found the resulting apocenter ratio is too small to be consistent with
observations.
We cannot formally rule out the proposition that
older debris could be reconciled with the observed apocenter ratio, 
perhaps with a more flexible model of the Milky Way potential.
We can only say that our limited numerical experiments did not produce any
successful models of this type.
As we will discuss below, it {\it is} possible that the
satellite actually experienced one or more earlier pericentric passages.
However, we infer from the spatial structure of the RR Lyraes 
(e.g., the low density in the inner trailing tail past apocenter)
that these earlier passages
were less disruptive to the stars than the last two, due to evolution of the satellite orbit and
internal structure, and thus formed more tenuous stellar stream components if any.
Such an earlier period of the satellite's history is still consistent with the material
near leading and trailing apocenter originating in the last two full pericentric passages.
Our single-component satellite models lack an enveloping dark halo---expected to be
stripped before most of the stars, and therefore relatively more important in this
earlier evolution phase---so we exclude it from consideration here.

In the following work, we therefore enforce the assumption that the 
the progenitor has experienced just three pericentric passages.
(In contrast, the \citetalias{law10} model had seven.)
The first results in an extended stream,
the second in a less extended stream that deviates more from
the orbit, and the third is the currently ongoing one.
(In our models, the Sagittarius dSph is consistently just past its latest pericentric passage.)
To be specific, we impose a restriction that the satellite experiences only two
apocentric passages since the beginning of the model integration,
making no other demands on the initial orbital phase.
(This setup actually allows the satellite to begin its evolution after the ``first''
pericenter just described.  This normally results in an underdeveloped initial stream
and thus such states are disfavored in our samples, but a few do occur still.
We have excluded such states from the plots below.)

We also enforce our interpretation of the stream substructure
in Section~\ref{sec.obs_structure}
when selecting model particles to compare to the observed distance data.
Specifically, we fit the trailing apocenter
distance points using only the older component liberated around the
first pericentric passage.  Similarly, we fit the distances around
leading apocenter with only the younger component liberated at the second passage.
These choices deliberately impose a particular general structure on the stream models, 
which as we will see is testable with future observations.
\subsection{Dynamical modeling}
\label{sec.dynamics}
Our dynamical models assume a single, spherical, hot, and non-rotating
component in the progenitor,
where the mass approximately follows the light. Of course, any
cosmologically-motivated model should include dark matter with a
more extended density profile than that of the stars.  We essentially
assume here that any extended dark matter component has been tidally stripped 
in pericentric passages predating the start of the simulation.
We also ignore possible internal structure in the angular momentum
distribution, as in the disky model of \citet{jorge10}.

The spray models are calculated using the recipe presented in \citet{fardal15},
where it was found to reproduce well the dynamical structure of tidal streams.
It includes a prescription for variable mass loss along the satellite orbit.
The combination of pulsed mass loss and strong tidal forces at pericenter
produce a series of streams originating at different pericentric passages,
consistent with \nbody\ results.
The particle orbits in these models
are calculated with the Python/C package {\tt galpy} \citep{bovy15}.
Although the recipe used here is successful in various respects,
a comparison of Figs.~\ref{fig.nbodyex} and \ref{fig.sprayex}
shows some differences between the spray models and the more reliable \nbody\ models.
In particular, the S-shaped tracks near the satellite in the \nbody\ stream
are not modeled correctly in the spray models.
This is by design, as we omit the force from the satellite 
to ensure the particles escape immediately upon release.   
For a similar reason, encounters between the satellite and its own extended streams
are ignored.
Despite these and some other less obvious deviations from the \nbody\ results,
the spray models are extremely useful because of their lower computational
cost compared to \nbody\ models, and because our greater knowledge of the particle
properties (such as ejection times) in these models can be used directly
in fitting the stream.

The spray models also make it feasible to include a crude treatment of dynamical friction,
without the computational burden that \nbody\ simulations of live host components would impose.
We use the standard Chandrasekhar formalism to compute the decay of the satellite orbit.
Orbits of released particles are instead computed without dynamical friction.
This approach is valid if particles escape quickly enough from the vicinity of the
satellite to stop feeling the dynamical friction perturbation over most of their
orbits.  Within the context of the Chandrasekhar approach we expect most of the
frictional force to be localized within a few kpc of the satellite, whereas the
particles near apocenter have moved 50--100~kpc away, so this simplification 
may not be too unreasonable.
We assume a Maxwellian velocity distribution in the host.
We compute the required velocity dispersions
using the Jeans equation with a ``sphericalized'' version of the host
potential under consideration.  We set the Coulomb logarithm to a
fixed $\ln \Lambda_c = 10$.  This is higher than the value of $\tsim 3$
we would obtain from the formalism of \citet{petts16}.  
One motivation for choosing such a high $\ln \Lambda_c$ is to bracket the behavior
with our no-dynamical-friction models.
Also, mass from a dark halo component
could enhance the dynamical friction even after being
formally unbound from the satellite \citep{fujii06}.

Our dynamical friction treatment thus involves many approximations.
A correct treatment of dynamical friction necessarily involves
treatment of the resonant nature of the interaction with the halo,
which may be difficult to treat correctly even in live-host \nbody\ simulations,
as well as folding in the highly uncertain mass loss history of the progenitor
starting from first infall.
Our main goal here is to illustrate the qualitative nature of the effects that
dynamical friction has upon the stream models.  As the observational situation
improves, more accurate models of the dynamical friction will probably become necessary.

Our \nbody\ models are performed with the code {\tt PKDGRAV} \citep{stadel01thesis}.  We
initialize the satellite as a single \citet{king66} $W=|\Phi(0)|/\sigma^2 = 3$ model,
consistent with the
assumed parameterization of mass loss in the spray model.  
We use only rigid Milky Way potentials for these runs,
since initializing and running models with live hosts is significantly
more expensive.  Thus all of our results with dynamical friction will
be based on spray models with the treatment described above.

Both spray and \nbody\ models use satellites that are resolved with 12,000
particles for the total mass of the satellite.  In spray models where
not all of the Sagittarius dSph mass is stripped by the end of the simulation, the actual
number of particles used is lower in proportion to the actual amount stripped.
We have tested the results with higher numbers of particles.
The particle-induced noise in the likelihood function is diminished when using more particles,
but this is unimportant here since precise parameter distributions are not our goal.
Otherwise we found no significant differences in the results using either method.
\subsection{Satellite mass and structure}
\label{sec.satmass}
The initial mass of the Sagittarius galaxy can be constrained in multiple ways.
The most direct way is to add up the stars visible in the satellite and stream.
This yields $5$--$8 \times 10^8 \msun$ with about
30\% remaining in the satellite currently \citep{niederste10}.
At this mass, cosmology suggests the galaxy should be associated with
a large dark matter halo of $\log_{10} M_{vir} = 10$--$11$
(\citealp{purcell11,gibbons17}; \citetalias{dierickx17}).
However, such a halo is also expected to be quite extended and rapidly stripped
in the first few orbits, during which time the orbit also decays due to dynamical friction
and the satellite becomes more vulnerable to stripping due to its smaller mass.
In simulation of this process by \citep{gibbons17} including both stars and dark matter,
the satellite loses so much dark matter as of two radial periods ago
that it retains a dark mass of only $\ltrsim 10^8 \msun$ and is largely
stellar-dominated.  

Another way to measure the mass
is to use the width of the stream in phase space,
with line of sight velocity dispersion and angular width the most practical measures at present.
The velocity dispersion in the trailing stream has been measured at
$8 \pm 1 \kms$ \citep{monaco07} using 2MASS selected red giants, or
$14 \pm 1 \kms$ \citep{koposov13} using SDSS giants.
In a new analysis of SDSS data,
\citet{gibbons17} reconciled these values as a difference between
metal-rich and metal-poor components.
\citetalias{law10} found a trend between $\sigma_v$ and initial satellite mass,
using \nbody\ models where mass follows light. 
Combining this trend with the \citet{monaco07} velocity dispersion, they inferred an initial
Sagittarius dSph mass of $6.4 \times 10^8 \msun$.
While the \citetalias{law10} trend was obtained with a single orbital model,
we have found our \nbody\ simulations using different orbits still roughly agree.
The \nbody\ models of \citet{gibbons17} that include dark matter and stars as separate
components find the bound mass two orbital periods ago was in the range
$5$--$10 \times 10^8 \msun$, producing masses and stream velocity dispersions
consistent with the \citetalias{law10} trend.  

Another possible way to measure the Sagittarius dSph mass is to look for
its effect on the Milky Way disk \citep{purcell11,laporte18}.  This technique is very promising,
since wavy features reminiscent of the simulations have been found in the
MW disk \citep{xu15}.
However, the origin of these features is not yet confirmed and precise measurements
of the Sagittarius mass with this method are not yet possible.
Another possible technique as we discuss below is to measure the effect
of dynamical friction on the structure of the Sagittarius debris.  Perhaps
eventually all of these methods will agree on the mass of Sagittarius, but for now
there is considerable uncertainty.

We have not included any term that strongly constrains the mass when fitting
spray runs to observational data, so instead we will simply assume single
fixed values for each run.  
Taking into account the $\sigma_v$-mass trend of \citetalias{law10} and the slightly larger
dispersion found by \citet{gibbons17}, 
most of the runs will use a single mass which we set to $10^{9.1} \msun$.
In one run with dynamical friction and low Milky Way halo masses
(the {\it TF-DF} run described below), we found it necessary
to reduce the satellite mass to $10^{8.5} \msun$ to consistently obtain
the required number of radial oscillations
when evolving the orbit into the past.

We parameterize the satellite radial profiles in terms of the ratio $f_t$ of the King model's
outer radius to that of the initial tidal radius at apocenter
(calculated for an orbit without dynamical friction).
We have simply adopted a fixed value of $f_t = 0.8$ for all our models.
Given the similarity of the plausible orbits, this results in similar though
not identical mass loss histories in all runs.  
Typically the satellite preserves $\sim 40\%$ of its mass by the end of the spray or \nbody\ run.
This value is reasonable in view of the estimate of $\tsim 30\%$ from \citet{niederste10}.
\subsection{Models for the Milky Way potential}
\label{sec.potential}
We use two previously specified ``standard'' Galactic potentials without free parameters,
and several ``adjustable'' potential families where we allow the parameters to vary.
Our first standard potential, {\it galpy2014}, is based on
the default {\tt MWPotential2014} model included in the {\tt galpy} package \citep{bovy15}.
This model contains a spherical bulge, Miyamoto-Nagai disk, and a spherical NFW halo.
The Miyamoto-Nagai disk has scale length $a=3 \kpc$,
scale height $b=0.28 \kpc$, and mass $M_d = 6.8 \times 10^{10} \msun$.
The density of the NFW halo
is parameterized as
$\rho(r) = \rho_h x^{-1} (1+x)^{-2}$ with $x = r / a_h$ and $\rho_s = M_h / (4 \pi a_h^3)$.
This has a scale length $a_h = 16 \kpc$ and $M_h = 4.37 \times 10^{11} \msun$.
To speed up the potential evaluation, we have replaced the original bulge
form in the {\tt galpy} model with a Hernquist model with parameters
$M_b = 4.5 \times 10^9 \msun$ and $a_b = 0.442 \kpc$.
This substitution makes a negligible difference to the total mass profile beyond
$\tsim 10 \kpc$, the minimum radius probed by our stream debris.

The other standard potential is the best fit ``truncated-flat'' or {\it TF} potential 
from \citet{gibbons14}, which uses the form
$V_c^2(r) \equiv G M(<r) / r = V_0^2 [ 1 + (r/r_s)^2]^{-\alpha/2}$.
Thus the rotation curve behaves as $V_c^2 \propto r^{-\alpha}$ for $r \gg r_s$,
and positive $\alpha$ represents a falling rotation curve.
We set the parameters of this model from the
center of the distribution in Fig.~12 of \citet{gibbons14}:
$V_0 = 225 \kms$, $\alpha=0.55$, and $r_s=15 \kpc$.

The first adjustable family of gravitational potentials we use is a single power-law,
$\Phi_\mathit{1PL}(r | V_0, r_1, \alpha) = -\alpha^{-1} V_0^2 (r /r_1)^{-\alpha}$.
We keep the reference radius $r_1$ fixed and allow $V_0$ to vary.
In this model, $\alpha$ has the same meaning at large radius as in the previous model.

The next family implements an upward-bending power-law, 
$\Phi(r|V_0, \alpha_1, \alpha_2, r_1, r_s)
= f_0 \Phi_{1PL}(r|V_0, r_1, \alpha_1) + (1-f_0) \Phi_{1PL}(r|V_0, r_1, \alpha_2)$.
Here the inner and outer potential slopes are $\alpha_1$ and $\alpha_2$.
$f_0$ is specified in terms of a transition radius $r_s$
where the rotation curves from the two components cross, 
so that $f_0 = q_r / (1 + q_r)$ with $q_r = (r_s/r_1)^{\alpha_1 - \alpha_2}$.
We require $\alpha_1 > \alpha_2$, so that the potential slope steepens toward the center.
Certainly this model's behavior is quite unrealistic within the solar radius.
As we will see later, however, it appears to be useful in fitting the stream
which orbits at larger distances.

The third adjustable model, {\it BDH} for ``bulge-disk-halo'',
builds on the {\it galpy2014} model, but makes several modifications.
The disk mass is scaled by $f_d$ so that $M_d = 6.8 f_d \times 10^{10} \msun$.
The NFW halo scale radius is scaled by $f_L$ so that
$a_h = 16 f_L \kpc$.
The NFW halo scale mass is similarly scaled by $f_M$ so that
$M_h = 4.37 f_M \times 10^{11} \msun$.
Also, the disk is optionally converted into a spherical Hernquist model, with a
rotation curve at large radius similar
to the original disk.  Specifically, we
use the same mass in this ``disk'' as for the original disk component,
while setting the scale length
$a_d = 1.8 \kpc$.
We refer to this fully spherical version of the model as {\it BDH-sph}.

In the additional model families {\it BDH-qz} and {\it BDH-qyqz}, we alter the {\it BDH}
model by changing the NFW potential 
to make the {\em potential} contours (not the density contours)
ellipsoidal: $\Phi(r) \rightarrow \Phi(r_{\mathit{eff}})$ where
$r_{\mathit{eff}}^2 = x^2 + (y/q_y)^2 + (z/q_z)^2$.
The {\it BDH-qz} model allows $q_z$ to vary but keeps $q_y$ fixed at $1$.
In {\it BDH-qyqz} we also set $q_y$ to a fixed value of 1.1, which we found by trial and error
was useful to improve the out-of-plane behavior of the stream.
In this paper we focus on the in-plane quantities, so in
the interest of keeping the number of free parameters low we have not
allowed arbitrary rotation of the ellipsoidal potential axes.
The fixed alignment of these flattening axes is suggested by the results of \citetalias{law10},
who found an optimum alignment of the potential axes only $7 \degree$ away from the $x$/$y$/$z$ axes.
\subsection{Observational data}
\label{sec.obsdata}
In Section~\ref{sec.obs_structure} we described the fitting process 
leading to the distance estimates and formal uncertainties in Table~\ref{tab.distance}.
Our goal in this paper is not to obtain rigorous parameter estimates, 
but to understand the physics involved in reasonable stream models.
Furthermore, we have not taken into account
any systematic error from the RR Lyrae distance scale; nor have we
analyzed the simulations and the observations in a strictly equivalent manner.
Hence we inflate the formal uncertainties, first adding a floor to the relative
distance error and then scaling up the result by a constant factor.
These inflated uncertainties, also listed in Table~\ref{tab.distance}
as ``adopted MCMC error'', are the ones used to generate samples from parameter space.
To use the best-populated part of the observed and simulated streams,
we restrict the longitude range used in the leading stream to 
$260\degree < \Lambda < 320\degree$
and in the trailing stream to $175\degree < \Lambda < 225\degree$.
Recall that $\Lambda$ is defined using the coordinate system of 
\citet{majewski03} and \citetalias{law10}, where the stream travels
in the direction of negative $\Lambda$.

We also use the binned stream velocities from red giant branch stars
tabulated in \citet{belokurov14} when fitting the stream models.
These agree well with the velocities of M giants presented by \citetalias{law10}
in their region of overlap, but also extend the measurements into the region
around trailing apocenter.
Again, we inflate the formal uncertainties by imposing an error floor and a
constant scaling to obtain looser uncertainties used in the MCMC runs.
We truncate the points used in the fit to
similar longitude ranges as for the distance dataset.  
The velocity data is listed in Table~\ref{tab.velocity}.

There are several other observables we have omitted from this likelihood function,
including the distance to the Sagittarius dSph,
the proper motion both of the dSph and of the stream debris,
the stream latitude, and the velocity dispersion within the stream.
These observables will be examined in Section~\ref{sec.modelprops}.

We set the current Galactic coordinates of Sagittarius to fixed values of
$l_{Sgr} = 5.5689\degree$ and $b_{Sgr} = -14.1669\degree$.
The distance $d_{Sgr}$ is more uncertain and we allow it to vary.
For all models, we follow \citetalias{law10} in assuming that
the tangential motion of the Sagittarius dSph
points in the direction of longitude $\Lambda$,
so we need specify only the total tangential velocity $v_{tan,gsr}$.
We set the radial velocity in the
galactic standard of rest (GSR) frame to $v_{rad,gsr} = 171 \kms$.
We start the model by computing the orbit backwards from the current Sgr location for
an evolution time of $t_{ev}$, another free parameter, though one constrained
by our previously stated conditions on the number of orbits experienced
during the simulation.

We use the solar reference frame specified in \citet{sohn16}.
This assumes a solar radius of $R_{\sun} = 8.29 \kpc$, a position in the galactic
mid-plane, a local circular velocity of $V_c(R_{\sun}) = 239 \kms$ \citep[from][]{mcmillan11},
and velocities relative to the local standard of rest of
$v_X = 11.10 \kms$, $v_X = 12.24 \kms$, and $v_X = 7.25 \kms$ \citep[from][]{schoenrich10}.
\subsection{Selecting model parameters}
\label{sec.mcmc}
To select plausible model parameters, we have performed
Monte Carlo Markov Chain (MCMC) runs with the spray models.
We adopt uniform priors on the parameters, with sharp cutoffs
that in most cases do not constrain the sampled parameter values.
We assume Gaussian uncertainties so that the likelihood function
takes the standard $\chi^2$ form.

We construct smooth trends of distance versus longitude from our particle data
using kernel regression (Nadaraya-Watson smoothing), and compare the results to
the distance data points given in Table~\ref{tab.distance}.
As explained in Section~\ref{sec.assumptions}, we use only particles
from either the younger or older debris stream at leading
and trailing apocenter respectively.
We also use the stream velocities from B14 as listed in Table~\ref{tab.velocity} in a similar manner.
In contrast to the distance, we use all the model particles regardless
of ejection time when fitting to the velocities, as the two components
are not clearly distinguished in velocity space at present.

Our adopted uncertainties on the distances
and velocities probably exceed the true uncertainties in the data.
This allows us to produce sample models in rough accord with the data,
without fearing that systematic model biases or underestimated uncertainties
will lead to strong and erroneous inferences about the Galactic potential
and other aspects of the model.
However, this approach does mean that the dispersions of the parameters
and the physical properties of the models are probably overestimated,
and certainly should not be interpreted as the true statistical uncertainties.

Our MCMC runs are conducted starting from plausible initial parameter guesses,
using the DE-MCMC algorithm of \citet{braak06} within the statistical code
\textsc{BIE} \citep{weinberg12bie}.
Although accurate statistical inference is not our goal here, 
we do monitor the parameter and likelihood values in the chains and
perform a Gelman-Rubin test to assess convergence.
Typically runs of 300-400 steps with 24--48 chains are enough to generate a
converged sample of states within any one model family.
The plotted parameter samples were constructed by discarding the first half of
each run and then taking 30--100 random states, which is sparse enough 
to make the states nearly independent samples.

Good states from each of several five different bulge-disk-halo models are
given for reference in Table~\ref{tab.states}.  The listed model
states are those that maximimize the recorded likelihood value in our
MCMC runs.  The likelihood function contains random noise from the
particle realization, which somewhat randomizes the selected parameter
values.  The five different model families incorporate potentials
ranging from spherical to triaxial, and include one family with
dynamical friction activated.  We include enough information about the
orbital initial conditions and satellite structure to allow
replication of the model states if desired.  Only the first model
listed, Model~A, is discussed in the remainder this paper.

\section{REPRODUCING THE STRUCTURE OF THE STREAM}
\label{sec.modelphysics}
\subsection{Influence of the radial profile: tests in near-spherical potentials}
\label{sec.spherical}
\begin{figure}
\onecol{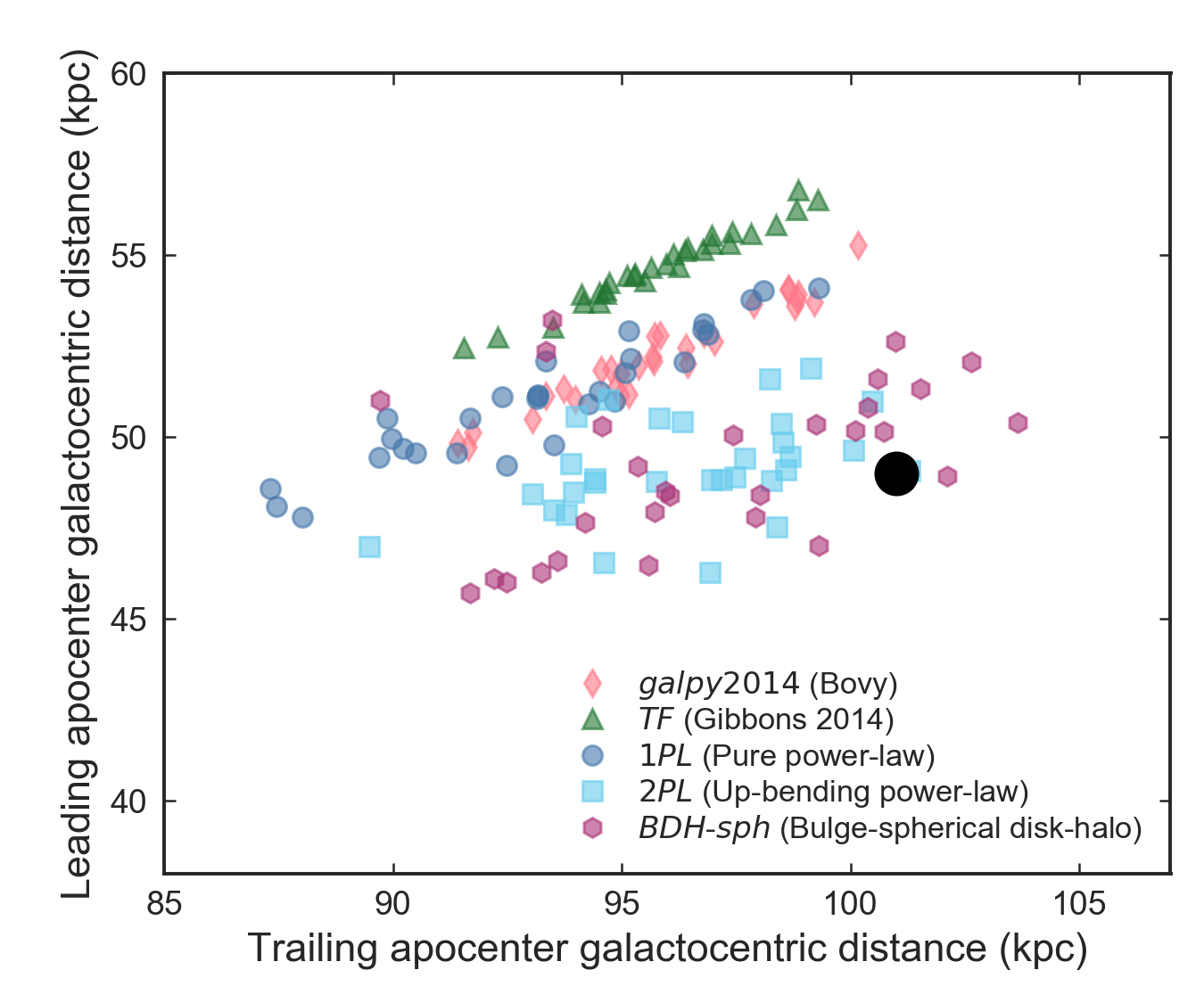}
\caption{
\label{fig.apoperi}
Comparison of the observed galactocentric radii at leading and trailing apocenter
(large black dot) to results from simulations (colored points).
Simulation points are for nearly-spherical model families without dynamical friction.
The values are measured for model particles in states subsampled from
the MCMC runs of the different model families, as indicated in the legend.
(Statistical uncertainties
on these radii are $\tsim 2\%$, while systematic error will affect both radii by the same factor.)
}
\end{figure}
We begin by examining models in spherical or nearly spherical potentials---i.e.,
we use the {\it galpy2014}, {\it TF}, {\it 1PL}, {\it 2PL}, and {\it BDH-sph} potential models.
We also exclude dynamical friction for now.
All of the sampled states with acceptable likelihood in these various models
have the same general appearance
as in Fig.~\ref{fig.sprayex}, but differ in the exact spatial and
velocity tracks followed by the streams.

The constant, ``standard'' models {\it galpy2014} and {\it TF} struggle
to produce as large an apocenter ratio as required, as does the adjustable family {\it 1PL}.
Fig.~\ref{fig.apoperi} illustrates this point.
We use subsamples of the model states to regenerate spray particle states, 
and estimate the leading and trailing apocenters
from the particle distribution
in each model using kernel regression.
The plots show the distribution of these values for the different model families.
The values obtained here for the {\it TF} model are consistent with those 
displayed in figure~8 of \citet{gibbons14}.  
While individual states in the constant-potential families can approach
{\em either} the leading or trailing stream value, they cannot reach both at once.
Consistent with this, the likelihood values for the standard models are
far worse than for the adjustable models (Table~\ref{tab.models}),
even though our likelihood function is quite tolerant by design.
The adjustable {\it 1PL} model achieves similarly poor results.
In contrast, the {\it 2PL} and {\it BDH-sph} families can reach the
observed apocenter values simultaneously.

\begin{figure}
\onecol{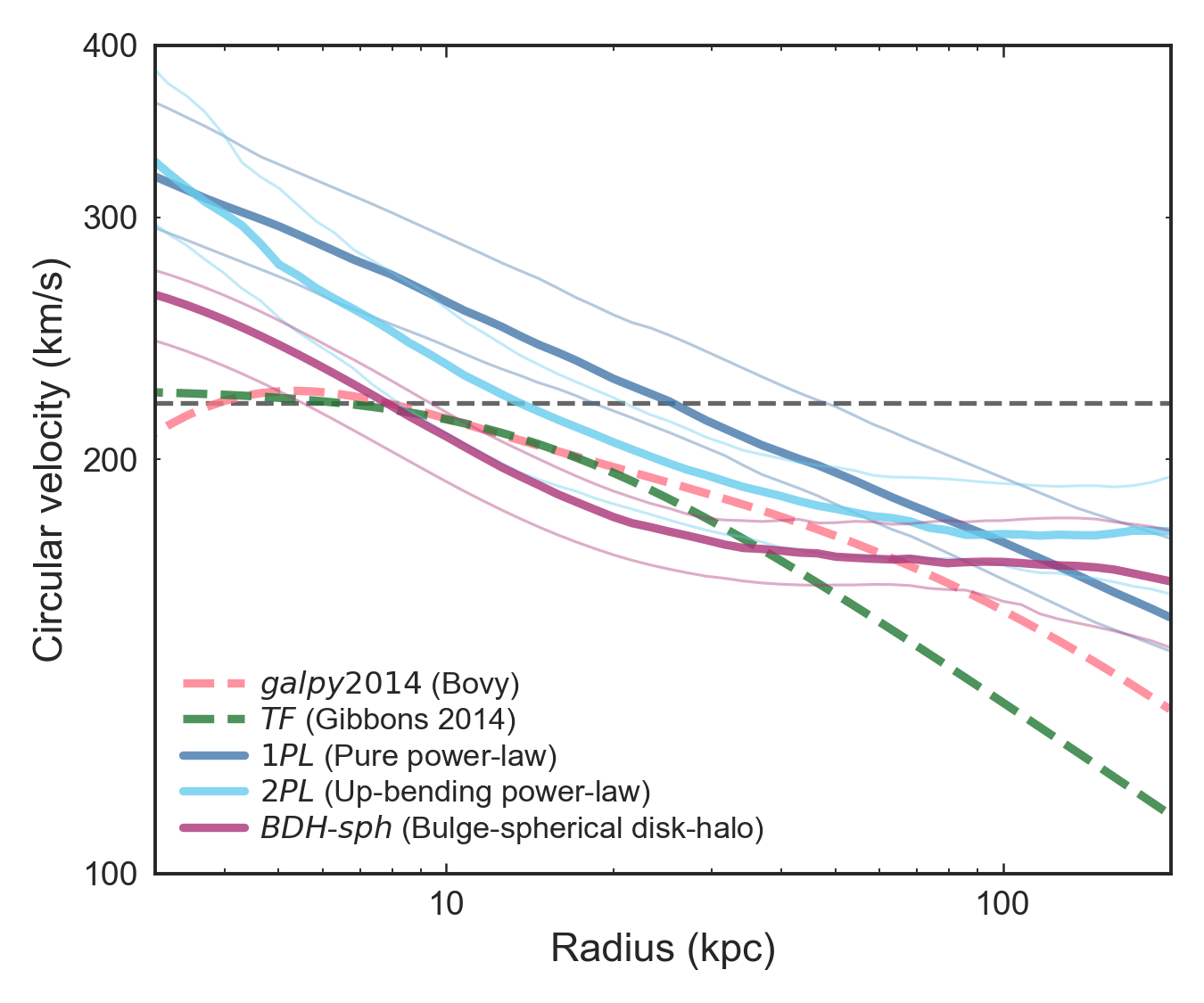}
\caption{
\label{fig.vccurves}
Rotation curves (circular velocities) in spray models as a function of Galactocenter radius.
Colored lines show results for samples from nearly-spherical model families without dynamical friction.
The solid curves show the median values measured in the sample,
and for the adjustable families the lighter curves show the 16--84\% range.
The dashed line shows $220 \kms$ for reference.
}
\end{figure}

Fig.~\ref{fig.vccurves} shows the circular velocity curves found by the various models,
as measured by their median and $16$--$84$\% ranges at each radius.
(Of course the standard {\it galpy2014} and {\it TF} potentials have no adjustable
parameters and no associated dispersions.)
The {\it 2PL} and {\it BDH-sph} models have similar upward-bending forms,
though with a slight roughly constant offset.
We have traced this offset to our adopted upper prior cutoff on the
disk mass scale $f_d$, which was intended to keep the rotation curve at small radius
at least somewhat reasonable.  
We performed another {\it BDH-sph} run after raising this limit (not plotted)
and obtained a mean rotation curve closer to that of {\it 2PL}.

\begin{figure}
\onecol{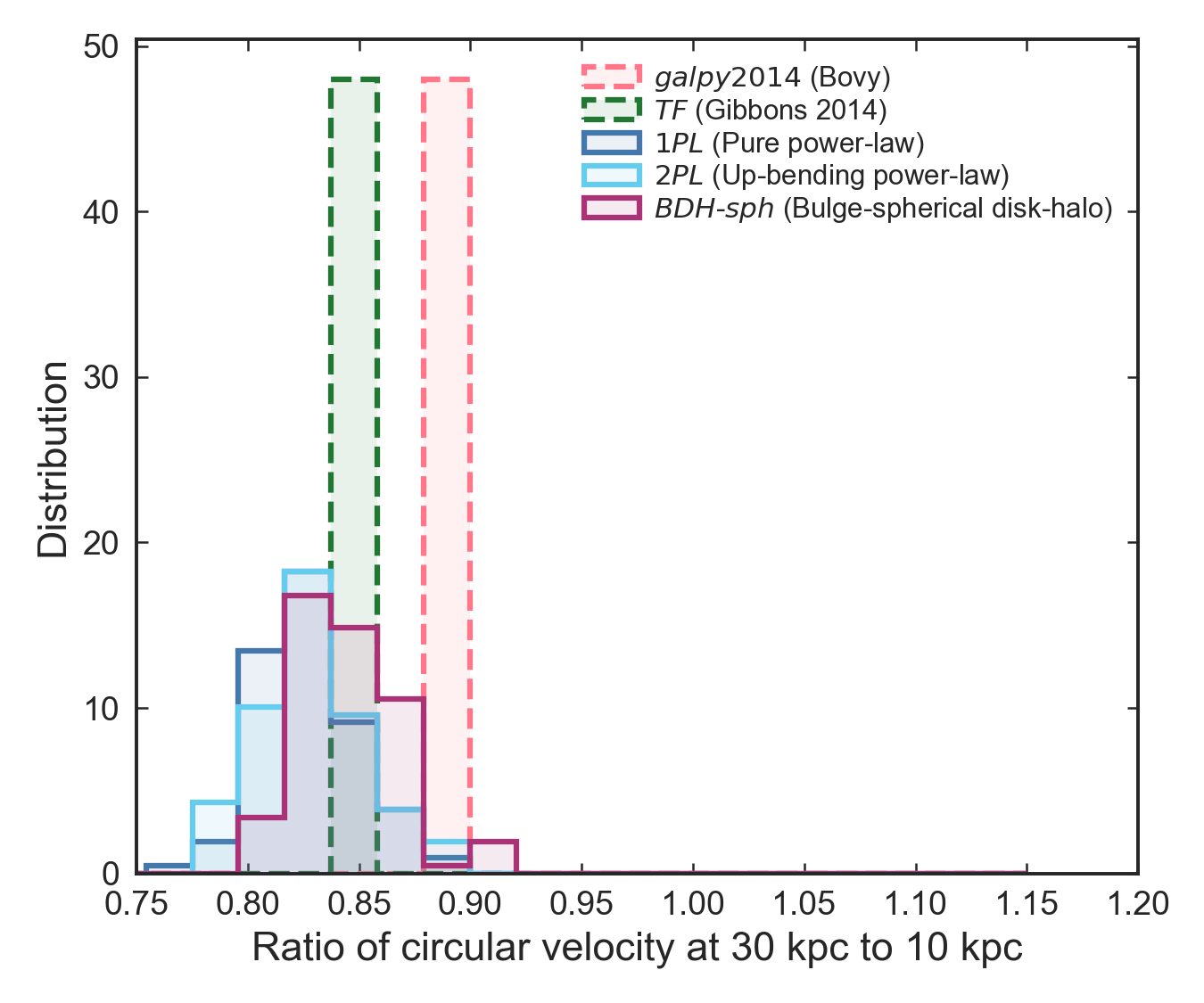}
\caption{
\label{fig.vcratio_inner}
Histogram of $V_c(30 \kpc) / V_c(10 \kpc)$,
the ratio of the circular velocity at 30 kpc to that at 10 kpc.
Each distribution shown is subsampled from one of our MCMC runs.
Here we include only nearly-spherical model families without dynamical friction.
}
\end{figure}

Fig.~\ref{fig.vcratio_inner} illustrates the constraints on the
inner halo circular velocity shape quantified by the
ratio of $V_c$ evaluated at 30~kpc to 10~kpc.  This ratio is fairly constant at around $\tsim 0.85$
in all the models.
\citet{gibbons14} argued that in order to fit the azimuthal position of the apocenters,
the halo rotation curve needed to fall off faster than
in the logarithmic halo used in the \citet{law05} and \citetalias{law10} models,
and this is seemingly borne out by our results.
The single or double power-law models could 
mimic the flat rotation curve of the \citetalias{law10} model,
but such states are poor fits and thus do not appear in our samples.

\begin{figure}
\onecol{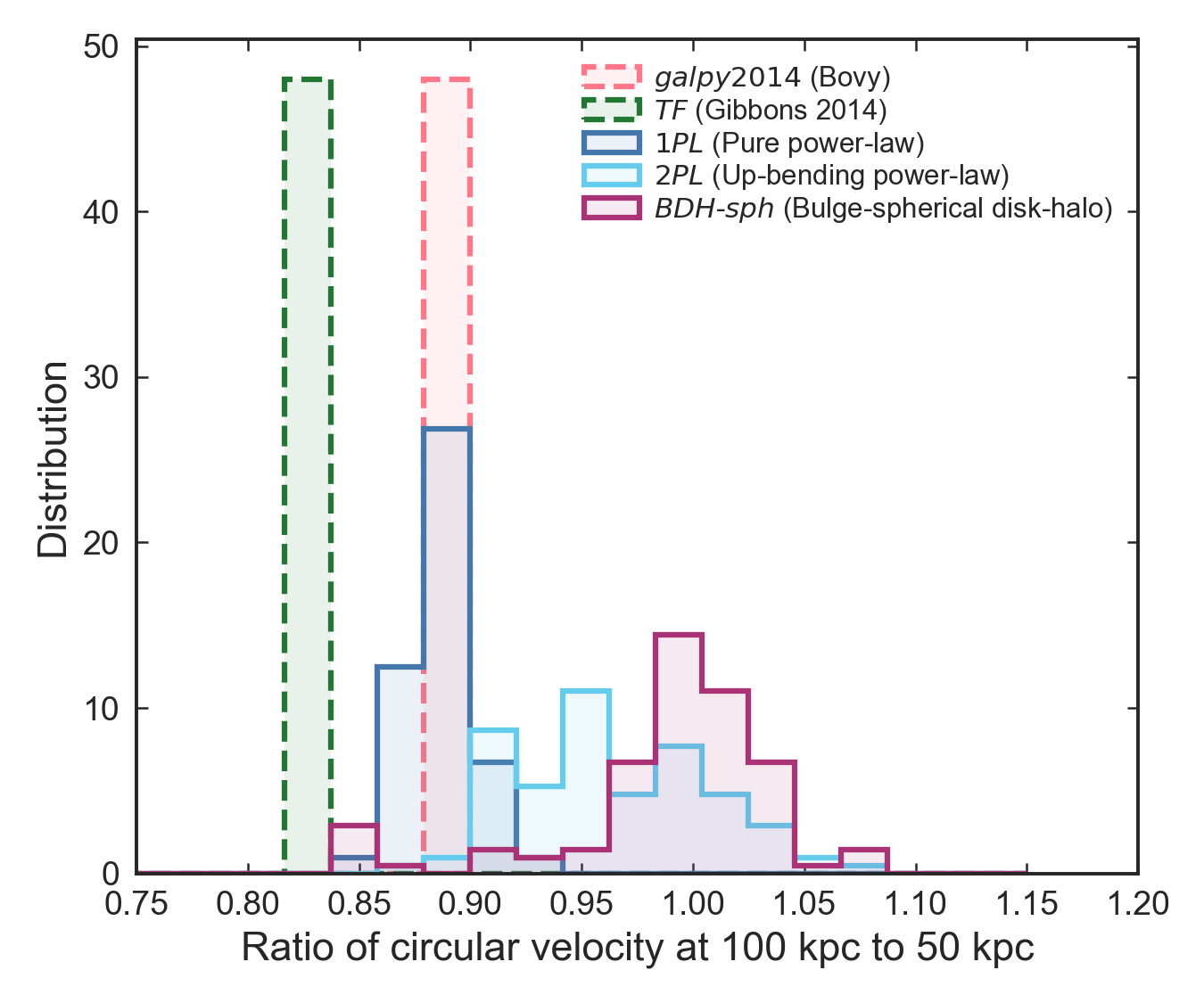}
\caption{
\label{fig.vcratio_outer}
As Fig.~\ref{fig.vcratio_inner}, but showing $V_c(100 \kpc) / V_c(50 \kpc)$,
the ratio of the circular velocity at 100 kpc to that at 50 kpc.
}
\end{figure}

Fig.~\ref{fig.vcratio_outer} 
shows the outer halo circular velocity shape measured by the
ratio of $V_c$ at 100 to 50 kpc.  In contrast to the previous one, this slope indicator
is markedly different between the different models.
For the better-fitting {\it 2PL} and {\it BDH-sph} models, the rotation curve,
instead of continuing to steepen, instead become shallower in a log-log
plot in the outer halo, only steepening past $\tsim 100 \kpc$ if at all.

At first this may seem counterintuitive: it takes less energy to lift stars
to $100 \kpc$ if the rotation curve falls off more steeply.
We must remember however that we are not investigating the maximum radius
at any time conditional on a given energy, but conditional on the stars being at
apocenter {\em now}.  E.g., for the old, curved trailing stream,
the stars must move out to $100 \kpc$,
then back to $\tsim 15 \kpc$, then out again,
all in the same time that the Sagittarius dSph
has completed two orbits of a smaller scale.
Just as a player who wishes to dribble a basketball higher and higher 
at a fixed frequency must exert stronger and stronger forces, 
the large estimate of the apocenter radius demands a strong halo force
in the vicinity of 100~kpc.

We can quantify this argument as follows.
Since stars on a highly radial orbit spend
the most time near apocenter, the orbital period is roughly
$T = k R_{apo} V_c^{-1}(R_{apo})$ where $k$ is a constant weakly dependent on the potential slope.
The rotation curve at leading and trailing apocenter are then related by 
\begin{equation}
\frac{V_{tr}}{V_{ld}} = \frac{R_{tr}}{R_{ld}} \frac{T_{ld}}{T_{tr}} \; .
  \label{eqn.vcratio_general}
\end{equation}
Here $V$, $R$, and $T$ refer to the circular velocity near apocenter,
apocenter radius, and orbital period, with subscripts denoting the
leading and trailing streams which are released near pericenter.
In the case of the Sagittarius stream, the young leading stream performs
1.5 orbital cycles of time $T_{ld}$ since release
in 1 Sagittarius dSph orbital period $T_0$, or a time 
$t_{ld} = 1 T_0 = 1.5 T_{ld}$, and observationally reaches $R_{ld} = 50 \kpc$ (galactocentric).
The old trailing stream instead
performs 1.5 orbital cycles in 2 Sagittarius dSph orbital periods,
or a time $t_{tr} = 2 T_0 = 1.5 T_{tr}$, and reaches
$R_{tr} = 100 \kpc$.
Thus for all stream models under consideration in this subsection,
\begin{equation}
\frac{V_{tr}}{V_{ld}} = \frac{R_{tr}}{R_{ld}} \frac{T_{ld}}{T_{tr}} = \frac{1}{2} \frac{R_{tr}}{R_{ld}} \; .
  \label{eqn.vcratio}
\end{equation}
For a stream actually matching the observed apocenter radii of 
$R_{tr} \approx 100 \kpc$, ${R_{ld}} \approx 50 \kpc$, 
\begin{equation}
\frac{V_{tr}}{V_{ld}} = \frac{1}{2} \frac{100 \kpc}{50 \kpc} = 1 \; .
\label{eqn.vcratio_obs}
\end{equation}
In other words, the rotation curve is flat from 50 to 100~kpc for a model
matching the stream.

\begin{figure}
\onecol{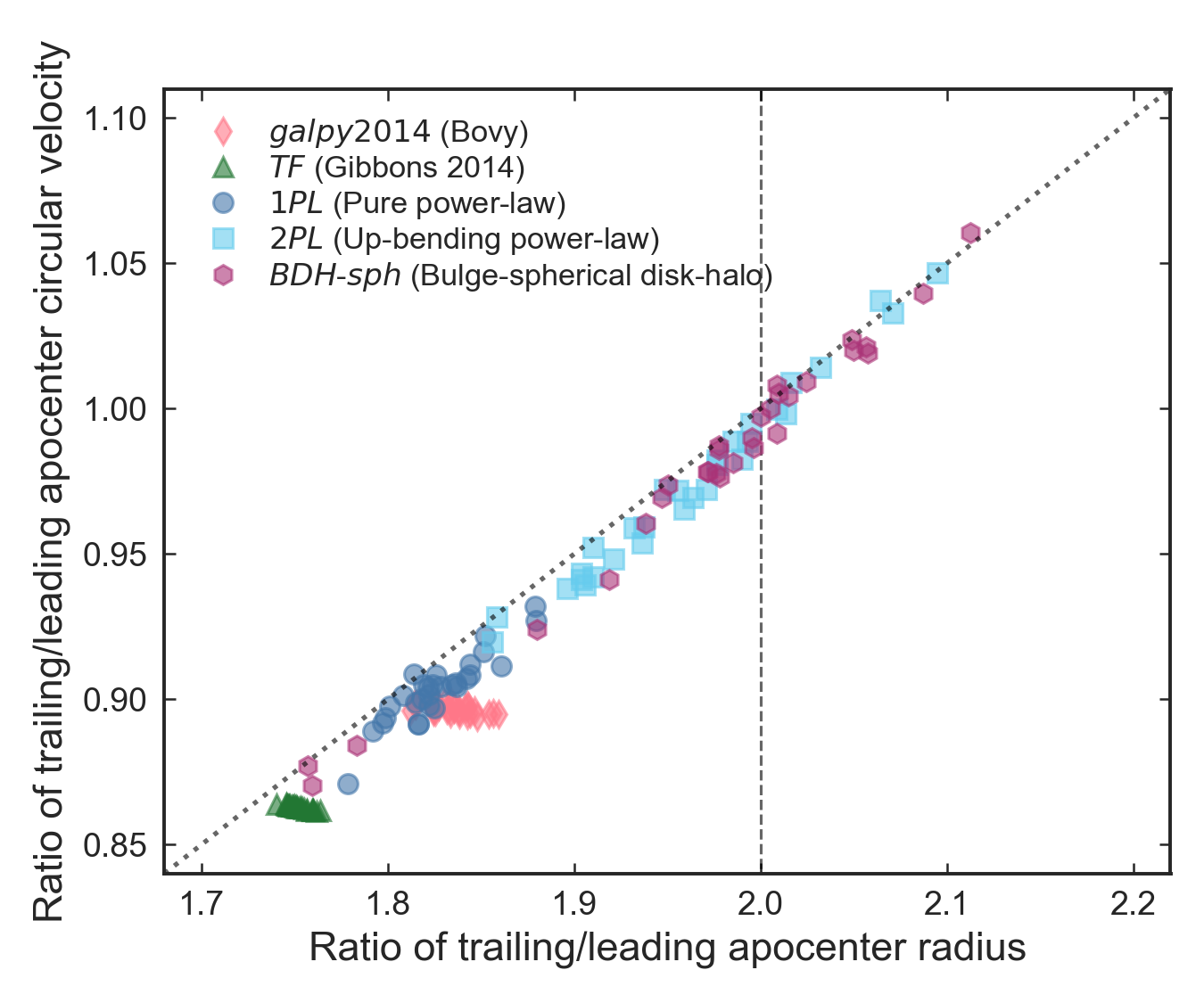}
\caption{
\label{fig.apoperivcratio}
Horizontal axis shows the ratio of the trailing and leading apocenter galactocentric
radii in spray models,
as compared with the ratio of the circular velocities at those radii on the vertical axis.
Model results are from nearly-spherical model families without dynamical friction.
The diagonal dotted line shows the prediction of Equation~\ref{eqn.vcratio}.
The vertical dashed line shows the observed apocenter ratio.
}
\end{figure}

Certainly this argument is only an approximation.  However,
Fig.~\ref{fig.apoperivcratio}
shows it holds to high accuracy for the current class of models
(near-spherical potentials and no dynamical friction).
The ratio of circular velocities at leading and trailing apocenter
is predicted almost exactly by Equation~\ref{eqn.vcratio}.
The only model with significant offsets
from the general trend is the not-quite-spherical {\it galpy2014}.  This suggests
that departures from sphericity can somewhat relax the tight relation, a point
we will examine further below.
The strong preference for nearly flat rotation curves between 50 and 100~kpc
is unexpected in standard galactic models,
but could suggest a more massive and extended dark halo than envisioned in those models.  
\subsection{Influence of dynamical friction}
\label{sec.dynfric}
\begin{figure}
\onecol{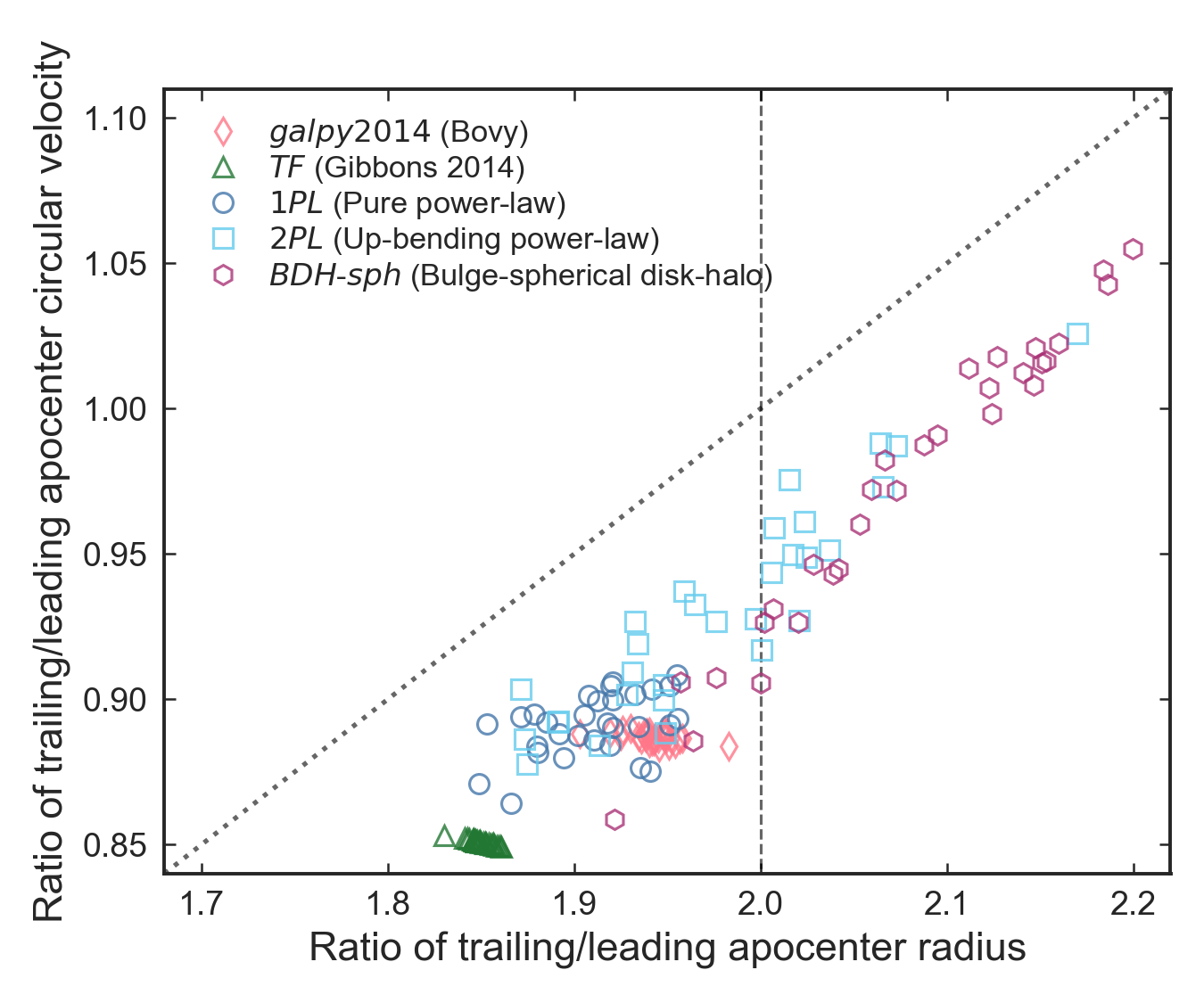}
\caption{
\label{fig.apoperivcratio_dfbad}
Apocentric quantity ratios as in Fig.~\ref{fig.apoperivcratio},
but now using model results from nearly-spherical model samples with dynamical friction included.
Reference lines are the same as the previous figure.
Note the significant shift and increased scatter compared with the previous figure.
}
\end{figure}
We now add dynamical friction to the satellite orbits in the manner discussed
in Section~\ref{sec.dynamics}, while continuing to use the same set of nearly-spherical potentials.
We measured the leading and trailing apocenters in a similar manner to the
previous runs.
The results are illustrated in Fig.~\ref{fig.apoperivcratio_dfbad}.
Clearly, the mean trend has shifted by a large amount compared to the 
results without dynamical friction in Fig.~\ref{fig.apoperivcratio},
and the scatter is greatly increased.

How can we understand these results?
The initial effect of dynamical friction is produced at the
first pericentric passage, when the satellite loses orbital energy.  
This means the stars released in the subsequent pericentric passage have lower
energy on average.  Alternatively, this can
be regarded as raising the energies and thus the orbital timescales
of the old stream relative to the young stream.  Since the orbital timescales and phase
are essentially determined by orbital energy, this displaces the old stream backwards
along its track, without greatly changing the location of this track.
Therefore the initial effect of dynamical friction is to shift the older stars
{\em along} the stream, not across the stream.  If dynamical friction were to
turn off after the first pericentric passage (due to high mass loss),
this would be the total effect.

This situation changes if the satellite also experiences significant dynamical friction
at its {\em second} pericentric passage.  In this case
the Sagittarius dSph no longer serves as a reliable clock; 
the orbital period of Sagittarius from second to third passage $T_2$ is
shorter than that from the first to the second $T_1$, and its current location
near pericenter now indicates an elapsed time of less than $2 T_1$
since the first disruptive encounter.  In this case, the ratio between apocentric
radii no longer is predicted to satisfy Equation~\ref{eqn.vcratio},
explaining the results in Fig.~\ref{fig.apoperivcratio_dfbad}.

\begin{figure}
\onecol{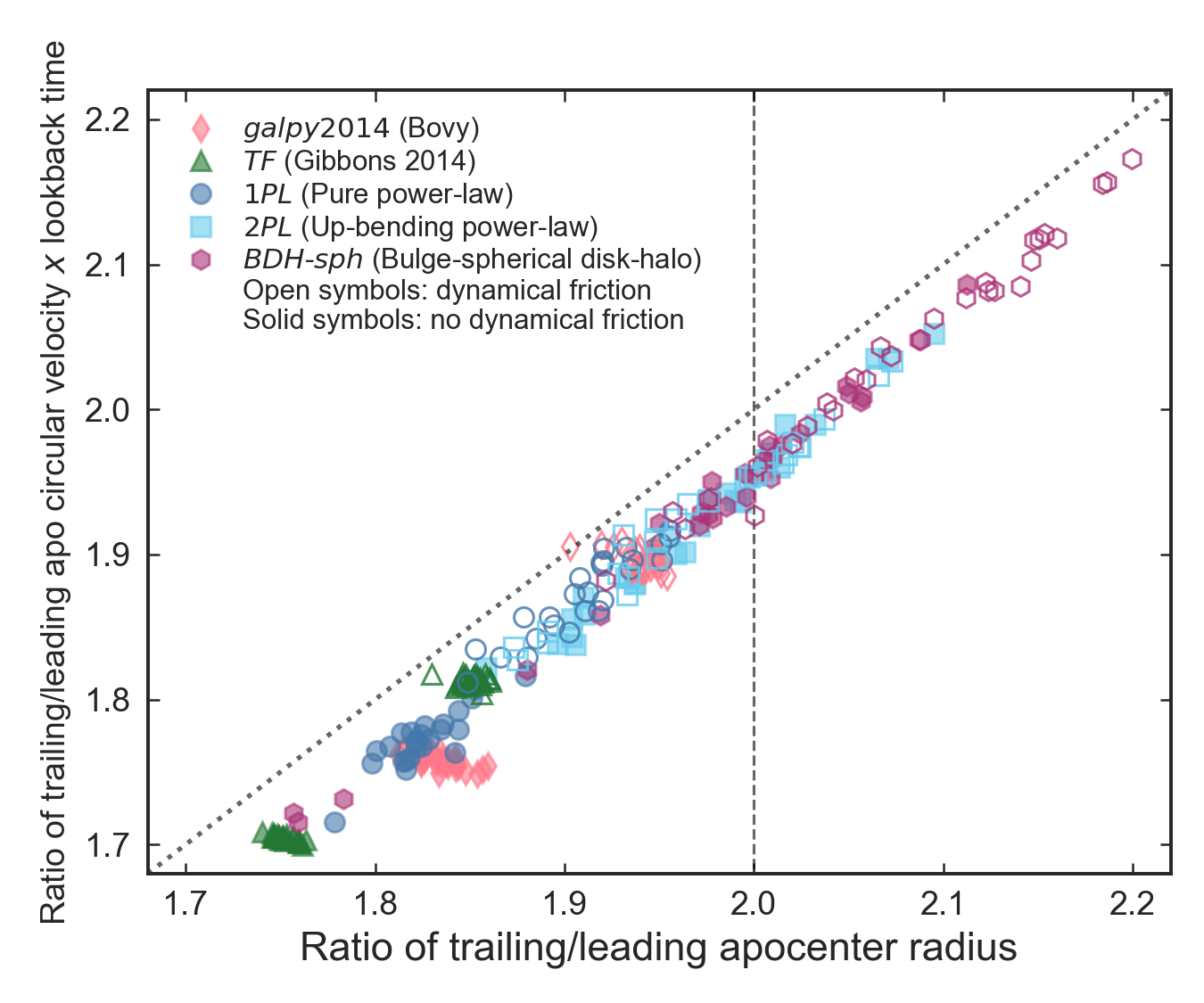}
\caption{
\label{fig.apoperivctimeratio}
Trailing to leading apocenter quantity ratios as Fig.~\ref{fig.apoperivcratio_dfbad},
but now the vertical axis is the ratio $Q_{VT}$ in Equation~\ref{eqn.vctratio},
i.e.\ the ratio of circular velocity times the lookback time to the relevant pericenter.
These lookback times are closely connected to the average particle ejection time
in the two stream components.
Open symbols show models with dynamical friction, and closed symbols those without.
The diagonal dotted line now shows the 1-to-1 line
(the prediction of Equation~\ref{eqn.vctratio}).
Note that the model points once again cluster into a tight relation,
though slightly offset from the prediction.
}
\end{figure}

We can still use the more general Equation~\ref{eqn.vcratio_general}.
Assuming the stars are released exactly at pericenter and
the Sagittarius dSph is also
at pericenter, this evaluates to
\begin{equation}
  Q_{VT} \equiv \frac{V_{tr}}{V_{ld}} \frac{t_{tr}}{t_{ld}} 
  = \frac{V_{tr}}{V_{ld}} \frac{T_1 + T_2}{T_2}
  = \frac{R_{tr}}{R_{ld}}  
  \label{eqn.vctratio}
\end{equation}
or for a model satisfying the observed stream apocenters
\begin{equation}
Q_{VT} = \frac{V_{tr}}{V_{ld}} \frac{T_1 + T_2}{T_2} \approx 2 \; .
  \label{eqn.vctratio_obs}
\end{equation}
Resampling the states of our various model families as before,
we measured the apocentric radii and rotation velocity at these radii,
and computed the lookback times to the first and second pericenter from the progenitor orbit,
taking them to be $t_{tr}$ and $t_{ld}$ respectively.  
Fig.~\ref{fig.apoperivctimeratio} compares the prediction of Equation~\ref{eqn.vctratio}
with the measurements of the
model samples, including both DF and non-DF runs.
The altered timescale factor in Equation~\ref{eqn.vctratio}
restores the tightness of the relation exhibited
by the no-DF runs in Fig.~\ref{fig.apoperivcratio},
though with an offset of about 2\%.
The offset is probably explained by the fact
the peak release time for the relevant particles is slightly {\em after} pericenter, 
rather than exactly at pericenter as assumed for simplicity here.

\begin{figure}
\onecol{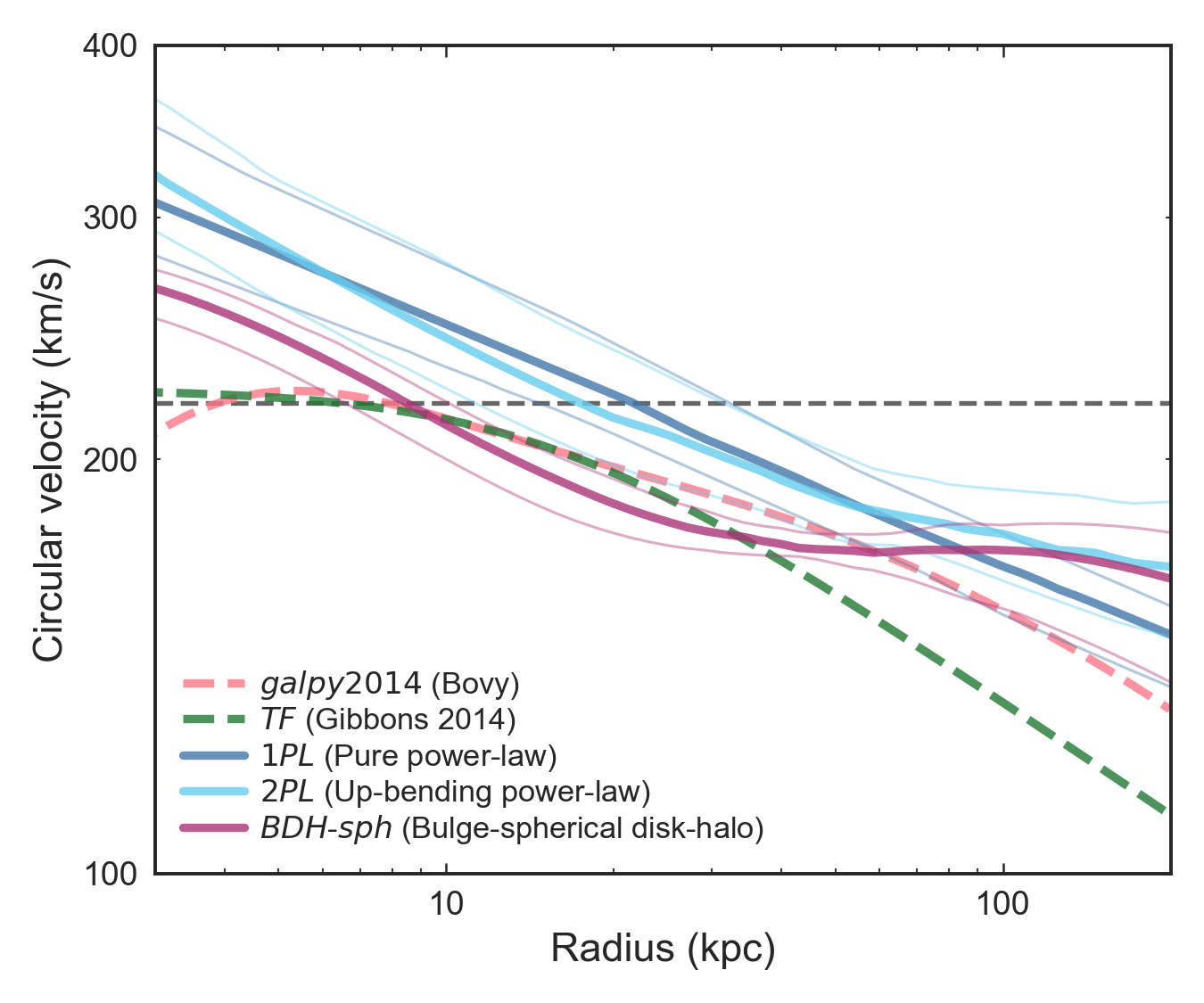}
\caption{
\label{fig.vccurves_df}
Rotation curves as in Fig.~\ref{fig.vccurves}, but for the near-spherical model
families with dynamical friction included.  Note the weaker upward bend in
the {\it 2PL} model in particular.
}
\end{figure}

Thus, we find dynamical friction weakens the previous conclusion about
the outer slope of the rotation curve, and allows somewhat steeper falloffs
for the fixed observational ratio of apocentric radii.
Table~\ref{tab.models} shows the standard models remain poorer fits to
the data than the {\it 2PL} or {\it BDH-sph} models, but the performance
gap is smaller than before.  This time the {\it 1PL} model is not too far behind
the best models.
The rotation curves from the model are shown in Fig.~\ref{fig.vccurves_df}.
While indeed less flattened at large radius, they do not qualitatively change
the picture from Fig.~\ref{fig.vccurves}.

Unfortunately, the 
tight relation in Fig.~\ref{fig.apoperivctimeratio} 
involves an 
unobservable ratio of timescales, which limits its use as a measure of
the outer halo rotation curve.
As already discussed, a theoretical estimate of the strength of dynamical
friction is laden with uncertainties.
However, it may be possible to constrain the effect of dynamical friction
from other observable quantities.  We consider one such method here.

We argued above that the bimodal appearance of the stream
at leading apocenter is likely a product of the older and younger stream
components both being present in this region.  Under this assumption
we can apply a similar argument as for the leading and trailing apocenter.
We assume here the circular velocity does not change much between
the inner and outer leading apocenter. In fact it usually changes by
$<2$\% in our models, because the difference in distance is small
and circular velocity is fairly flat.

The old stream at leading apocenter takes 2 progenitor periods
to complete 2.5 radial orbits, so its period is $T_{old} = (4/5) T_0$.
The young stream takes 1 progenitor period
to complete 1.5 radial orbits, so its period is $T_{young} = (2/3) T_0$.
With no dynamical friction, we then find
\begin{equation}
\frac{R_{old}}{R_{young}} \approx \frac{T_{old}}{T_{young}}
 = \frac{(4/5) T_0}{(2/3) T_0} = 1.2 \; .
  \label{eqn.split_nodf}
\end{equation}
In the case that dynamical friction changes the orbital period of
the Sagittarius dSph 
from $T_1$ to $T_2$, we instead find
\begin{equation}
R_{old} / R_{young} \approx 0.6 \frac{T_1 + T_2}{T_2} \; .  
  \label{eqn.split}
\end{equation}

\begin{figure}
\onecol{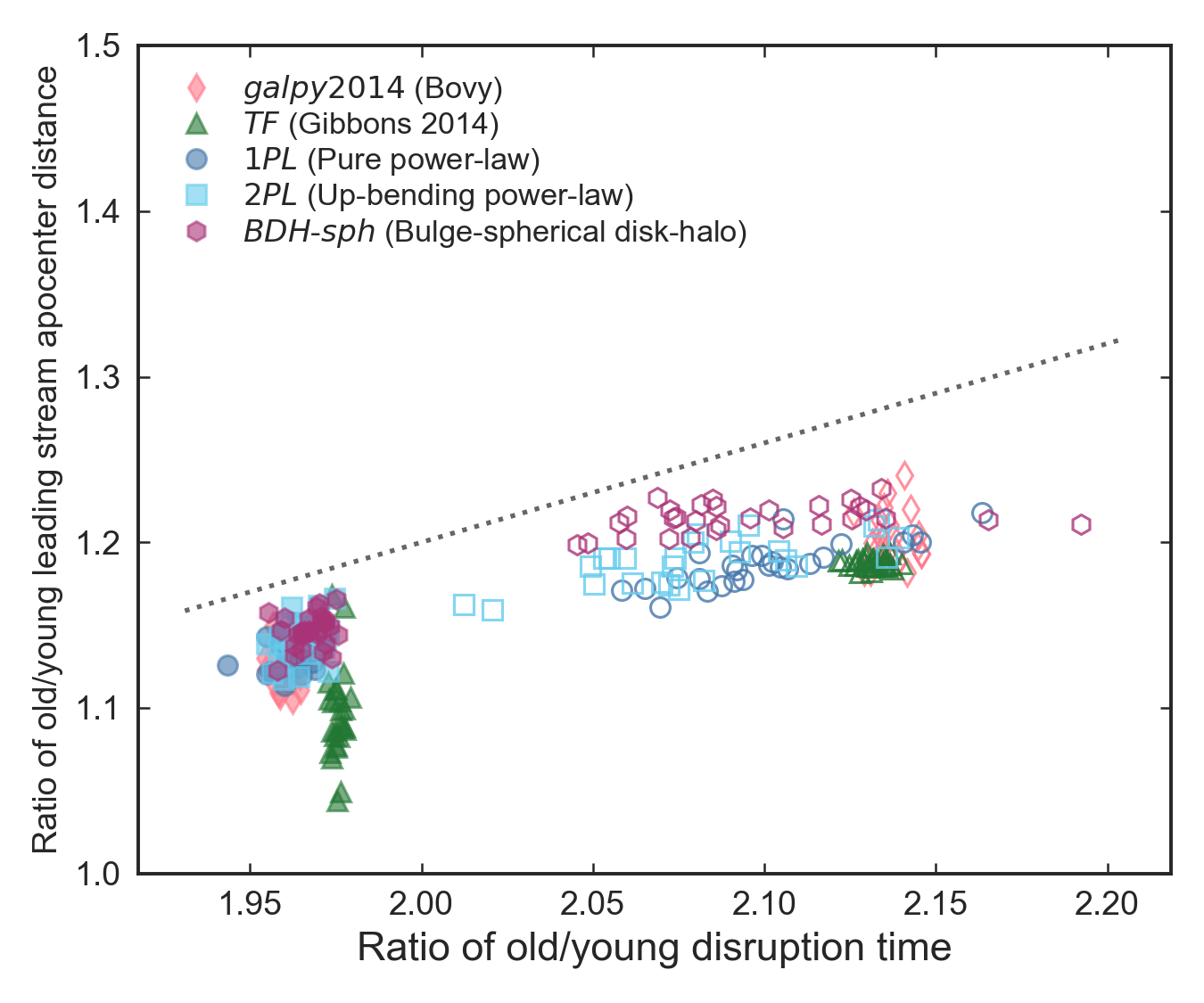}
\caption{
\label{fig.splitvstime}
Ratio of the galactocentric distance in the older and younger components at
apocenter of the leading stream, plotted versus the ratio of lookback times
to the first and second pericentric passages.
Values are measured in spray models with nearly spherical potential families.
Open symbols show models with dynamical friction, and closed symbols those without.
The diagonal dotted line shows the prediction of Equation~\ref{eqn.split}.
The substructure in the Pan-STARRS1 RR Lyrae suggests a ratio of component
radii $\approx 1.2$, although both the numerical value and even our
qualitative interpretation of the observed substructure are somewhat tentative at present.
}
\end{figure}

We have measured the leading apocenters in old and young components in our ensemble of runs,
along with the timescales $T_1$ and $T_2$.
The results are shown in Fig.~\ref{fig.splitvstime}.
The absolute calibration of our relation is
off by about 4\%, but the slope of the trend is quite good.
The increased split between components in runs with dynamical friction 
is easily apparent from visual inspection of plots like Figure~\ref{fig.sprayex}.

This plot suggests that we can constrain the ratio of 
the last two Sagittarius dSph orbital periods $T_1$ and $T_2$
from the separation of the two leading apocenters,
and thereby determine the overall effect of dynamical
friction on the stream.  From our two-component fits to the leading
stream region, we find the primary and ``fluff'' components
have a ratio of galactocentric radii of about 1.19.
This value is at least roughly consistent with
most of the models in Fig.~\ref{fig.splitvstime},
but agrees better with the models including dynamical friction.
This interpretation also disfavors values of 
the orbital timescale ratio $T_1 / T_2$ much
larger found than in the models here,
and thus implies a relatively weak effect of dynamical friction.

Of course, it is not yet certain that we are correctly interpeting the
fluffy component at leading apocenter.
An alternative interpretation is that stars in the old stream
are instead piling up at the outer Virgo overdensity of \citetalias{sesar17}.
This would imply a significantly larger ratio of timescales and thus
a much stronger effect of dynamical friction.  
Of course, it is also possible that {\em neither} component represents the older
component of the stream.  
Clearly these observed components deserve further study to
determine their motion and physical nature.

\subsection{Influence of non-spherical potentials}
\label{sec.aspherical}
We now turn to models that differ strongly from spherical symmetry.
These models are necessarily harder to interpret than those in the
previous section. 
However, they allow greater realism---after all, we know the Milky Way
has a disk component, and on cosmological grounds we expect the 
dark halo to be somewhat flattened and/or triaxial.
Also, we know from previous work that deviations from spherical
symmetry can strongly affect the Sagittarius progenitor orbit and
the shape of the stream.

We generated MCMC samples from these models:
{\it BDH}, {\it BDH-qz}, {\it BDH-qyqz}, {\it BDH-qyqz-DF}.
In other words, we begin by using a real disk (unlike the {\it BDH-sph} model
of the last section) but with the halo still spherical, then allow flattening along
the $z$ axis, then impose a flattening along the $y$ axis, then add dynamical friction.
The last three models have comparable likelihood values and are all formal
improvements over models considered in the previous sections.

We find the MCMC samples generally prefer prolate models ($q_z > 1$). This is consistent
with earlier work showing that prolate halos improved agreement with distances
and velocities in the leading stream \citep[e.g.][]{johnston05,law05},
and these observables are indeed the drivers 
behind the improvement in our likelihood function.  Essentially, the prolate models
unbend the leading stream so that the returning portion no longer passes so close
or even interior to the Sun.
The flattening parameter from our {\it BDH-qz} sample is $q_z = 1.17 \pm 0.10$.

Of course, earlier work also 
shows that prolate halos move the leading stream to positive values of latitude $B$,
contrary to observations that show negative latitudes \citep[e.g.][]{helmi04}.  
Changing $q_y$ from 1 to 1.1 as in the {\it BDH-qyqz} run
more or less cancels this offset, restoring
latitudes near zero in the leading stream without much effect on
the other parameters. In particular the vertical
flattening is almost unchanged at $q_z = 1.15 \pm 0.09$.
The \citetalias{law10} model used a different functional form for the potential,
but it is still interesting that they also preferred flattening parameters
$>1$ along the $y$ and $z$ axes and nearly equal to each other.  While
they allowed rotation of the principal axes in the disk plane, their
preferred orientation was rotated by a mere $7 \degree$ from the $x$ and $y$
aes.  Their preferred solution with $q_{y'} = 1.38$ and $q_z=1.36$ is so flattened
along the $x$-axis as to almost require negative densities in some
regions.  However, their figure~5 indicates a strong degeneracy along
the line $q_1 \approx q_z$, meaning the smaller flattening values we
prefer here are not strongly disfavored compared to their best fit.

However, making the models prolate also affects the ratio of apocenters.
This ratio {\em decreases} for a fixed potential
as the models become more prolate, which
makes it even harder to fit the apocenter ratio.
The main reason is that by coincidence the orbital lobes turn by
roughly $270 \degree$ per cycle,
in a plane nearly perpendicular to the disk plane.
Thus the orbital lobes more or less alternate between moving in the $z$
direction and moving in the disk plane.  
For prolate potentials, the radial loops pointing in the $z$
direction are elongated compared to those in the plane.
The circular velocity ($V_c = (d\Phi/d\ln r)^{1/2}$) at apocenter is
in contrast nearly unaffected.
Stars in the leading loop have longer periods relative to the last
full orbital period of the progenitor as $q_z$ increases,
while the opposite is true for the trailing loop.
This reduces the ratio of apocenters when the potential is fixed.
If we instead fix the ratio of apocenters, 
the circular velocity at trailing apocenter must be increased.

\begin{figure}
\onecol{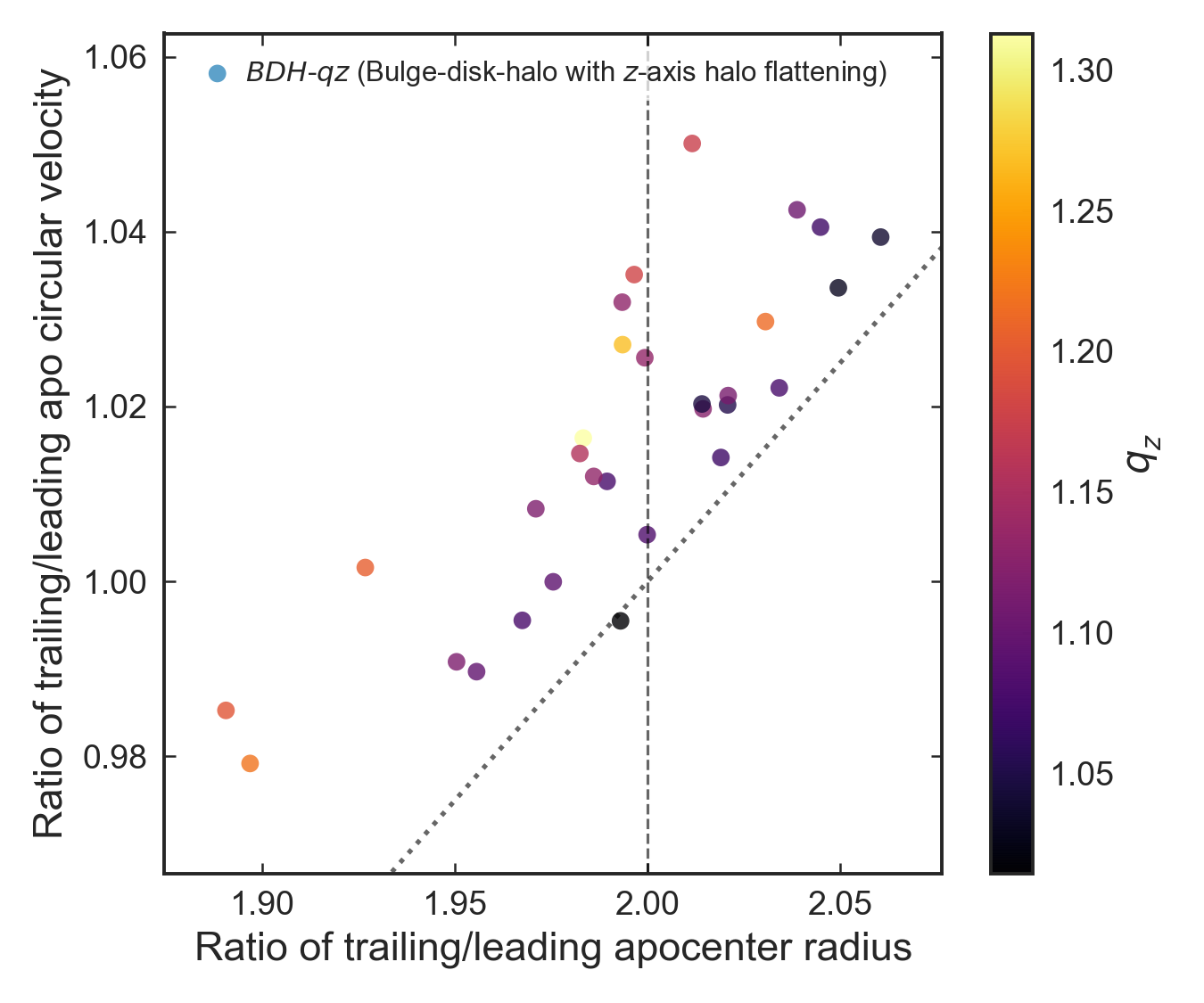}
\caption{
\label{fig.apoperivcratio_qz}
Ratios of galactocentric radii at apocenter and of circular velocities at those radii,
as for Fig.~\ref{fig.apoperivcratio}, but 
for the model family {\it BDH-qz} without dynamical friction.
The point color indicates the vertical flattening parameter $q_z$, 
where larger values are less oblate. All plotted points have prolate halos.
The diagonal dotted line shows the prediction of Equation~\ref{eqn.vcratio}.
The vertical dashed line shows the observed apocenter ratio.
}
\end{figure}

Indeed, we find generally larger mass and length scales for our NFW halo in the models
with variable $q_z$, making the rotation curve
slopes even less steeply falling at large radius than before.
Fig.~\ref{fig.apoperivcratio_qz} shows that these models no longer obey the tight
velocity-radius relation of Fig.~\ref{fig.apoperivcratio}, but lie above the previous relation,
in accord with the argument just given.
Note also the offset of the {\it galpy2014} models in the opposite direction
in Fig.~\ref{fig.apoperivcratio} demonstrates the same effect in reverse, 
since the potential in this model is mildly oblate due to the disk component.  
The degree of departure is correlated with the $q_z$, as shown by the color-coding.
At the same time, the changes are not particularly large, affecting the inferred ratio
of circular velocity in Fig.~\ref{fig.apoperivcratio_qz} by only a few percent.

In summary, aspherical halos can improve the
{\em overall} agreement of our models with observations of Sagittarius orbital-plane
quantities, much as described by earlier work.
However, the halo shape changes that produce better overall agreement actually
work {\em against} matching the large ratio of trailing and leading apocenter radii.
It is possible to imagine complicated potentials that vary from prolate within $50 \kpc$
to strongly oblate at larger radius, but finding a physically plausible model 
that accomplishes this with a {\it galpy2014}-like rotation curve would seem difficult.
If valid, the inference of a slightly prolate halo
{\em strengthens} the demands for relatively flat outer halo rotation curves
and orbital timescale changes due to dynamical friction, reinforcing the
conclusions in Section~\ref{sec.spherical}--\ref{sec.dynfric}.
\section{MODEL IMPLICATIONS}
\label{sec.modelprops}
In this section we illustrate the typical behavior of the models with one
illustrative \nbody\ state, model A, whose parameters are listed in Table~\ref{tab.states}.
This state is selected as one of the better states in the {\it BDH-qyqz} sample.
We use this model family lacking dynamical friction so that we can
easily construct the \nbody\ version.
The agreement of spray and \nbody\ versions is good but imperfect, 
slightly worsening the agreement of the \nbody\ model with observations.
The halo shape here is triaxial, with a fixed $q_y = 1.1$ and a fitted $q_z = 1.11$.
The disk mass is higher than in the {\it galpy14} model, and the NFW model highly
extended, as we discuss later.

\begin{figure}
\onecol{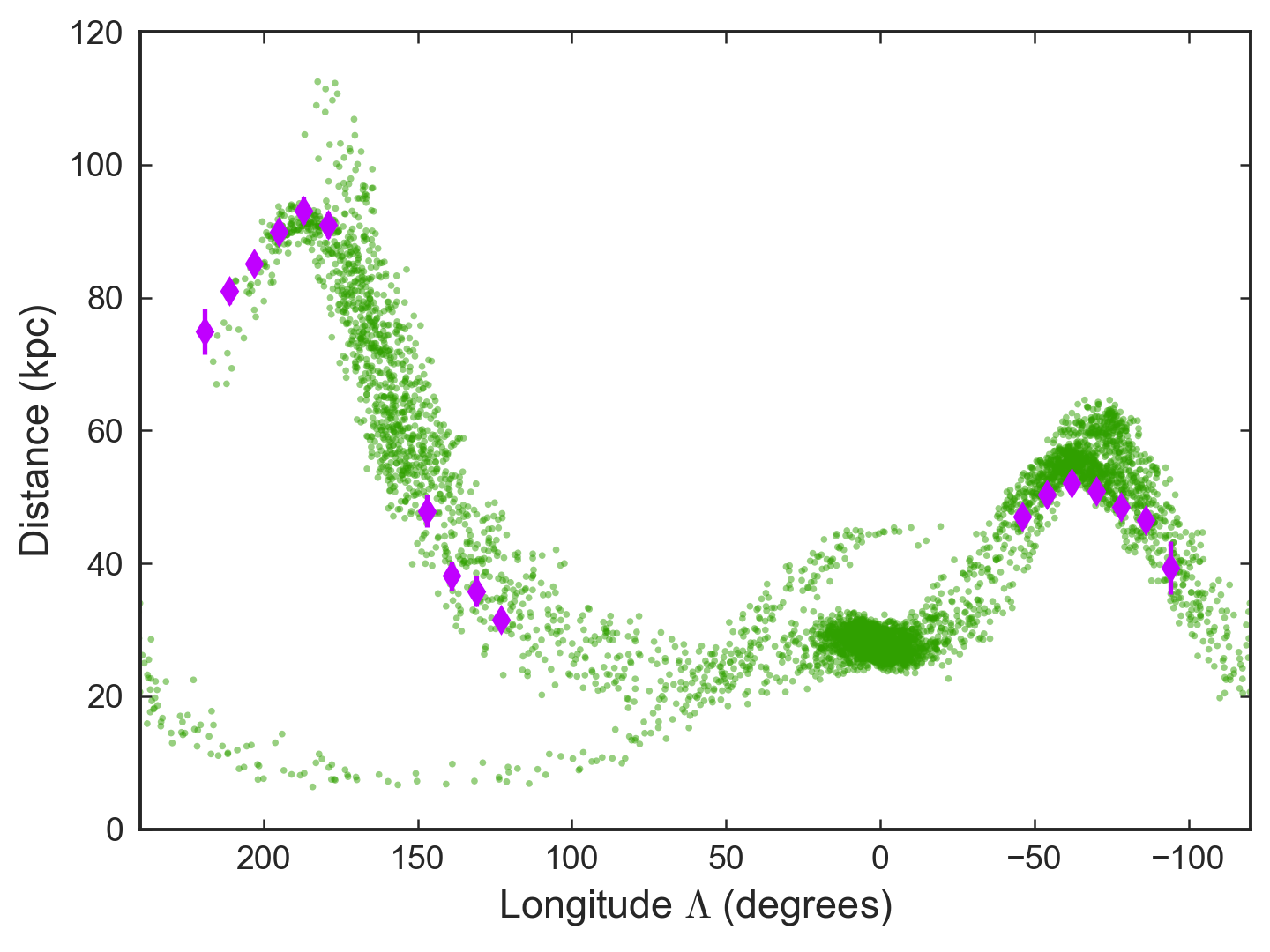}
\caption{
\label{fig.distance}
Line-of-sight distance to the stream stars versus stream longitude.  Model A particles
are shown by the green points.  The model
Sagittarius galaxy is the dense structure near $\Lambda = 0$. The main body of
the leading arm is on the right and the trailing arm is on the left.
The purple points show the mean distance estimates
obtained from the \citetalias{sesar17} RR Lyrae sample as stated in Table~\ref{tab.distance}.
The error bars show the statistical errors from the fit as stated in the table (not 
the ``Adopted MCMC error'' values which are significantly larger).  
}
\end{figure}

The behavior of the model in the Sagittarius orbital plane has already been shown
in Fig.~\ref{fig.nbodyex}.  In most respects the agreement is good.
The stream is not quite as extended at the extreme ends as the corresponding
spray run, which is a product of our spray models assuming
Gaussian distributions for simplicity \citep{fardal15}.
In the trailing stream at least, it also appears somewhat shorter
than the observed stream.  The young trailing stream may also be shifted
in azimuth compared to the corresponding component in the observations.

Fig.~\ref{fig.distance} shows the distance to the stream particles as
a function of stream longitude.  Sagittarius itself is visible as an elongated
dense structure at $\Lambda = 0$.  In this model, the agreement with the
leading and southern trailing stream points is mostly quite good.
A distinct ``fluff'' component at the leading apocenter and
a split in the stream at trailing apocenter are both visible.
In this model the leading stream continues to wrap around and overlaps with
the trailing stream for over $180\degree$ in azimuth.
The extent of this leading wrap is highly parameter-dependent.
While there are a few possible detections of this wrapped leading stream
over the range $0\degree < \Lambda \ltrsim 180\degree$
\citep{piladiez14,sohn15,hernitschek17}, much larger kinematic or proper motion
studies would be useful to reliably constrain the models.
The old trailing stream in this model nearly fades out 
for $\Lambda > 230 \degree$.  \citetalias{sesar17} and \citet{hernitschek17}
show the RR Lyraes in the stream probably continue to $\Lambda = 260 \degree$
and perhaps further, though the density in this extreme trailing tail is low
making its existence uncertain.
This apparent difference could be caused by stripping during earlier phases of the
Sgr dSph than we have modeled, or by in changes in the potential or orbit.

\begin{figure}
\onecol{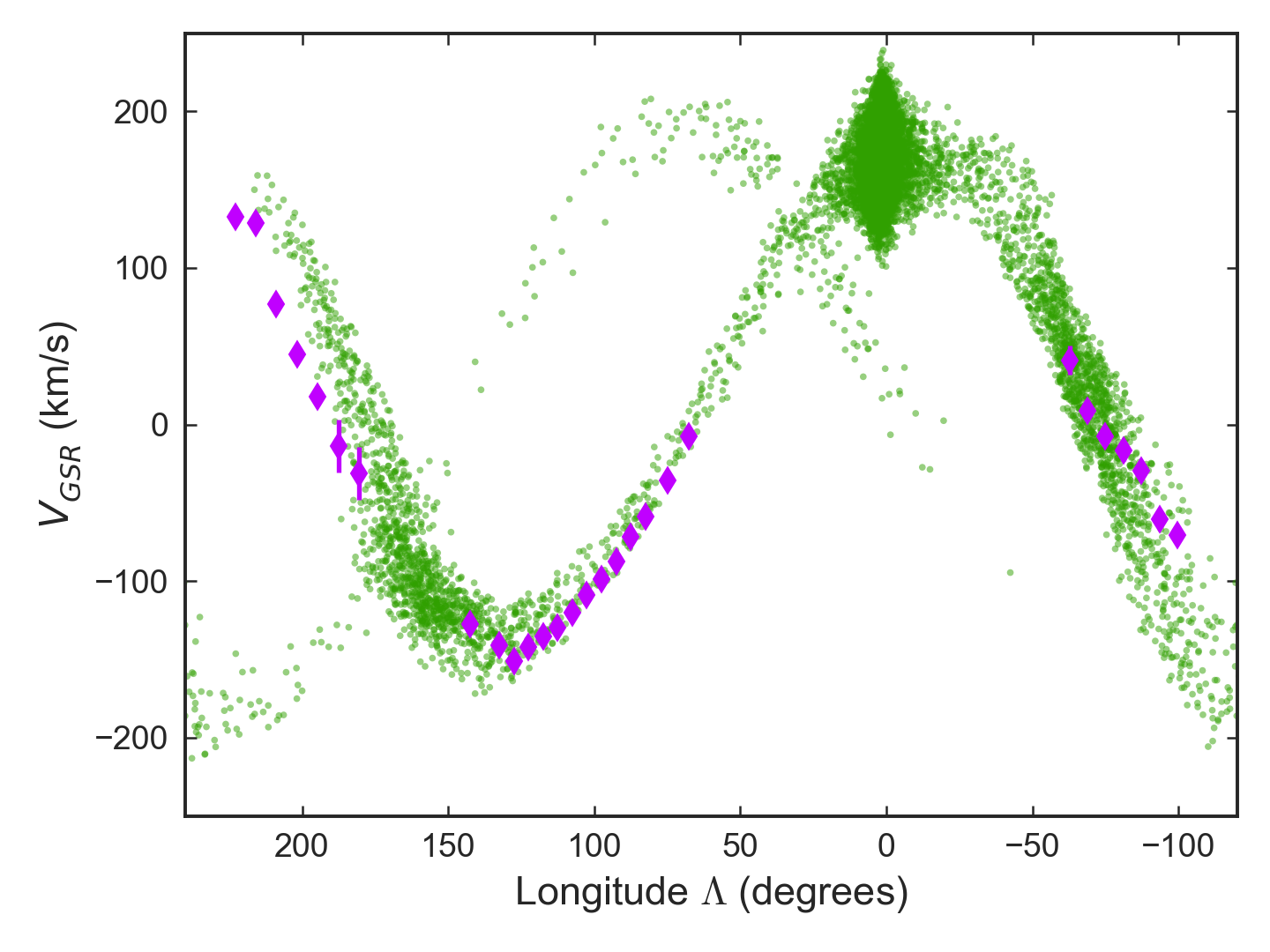}
\caption{
\label{fig.vel}
Galactocentric velocity of the stream stars versus stream longitude, analogous to
Fig.~\ref{fig.distance}.
Purple points show the observations of \citet{belokurov14} in our selected test regions
which are also supplied in Table~\ref{tab.velocity}.
}
\end{figure}

Fig.~\ref{fig.vel} shows the GSR radial velocity of the stream particles
along with the set of observations of \citet{belokurov14} we used in fitting.
Again, the agreement is reasonably good in the leading and southern trailing streams.
The leading stream is only matched this well in models with strong departures from sphericity.
In the distant trailing stream, there is a visible offset from the observational points
even before the simulation points die off.  This offset is also seen in spray runs,
where the model points do extend along the longitude range of the observations.
In our models there is often
a slight inconsistency between the distance and velocity at a given
longitude in the trailing stream.  The best-fit spray model appears somewhat
overprecessed in the trailing lobe compared to the spatial observations, whereas
they appear underprecessed compared to the velocity observations.
Both offsets are $\tsim 10 \degree$.
We speculate that the relative balance of old and young components in this region,
which is poorly constrained at present, may influence these offsets from the observations.
We postpone further examination of this issue to future work.

By comparing to the \citetalias{sesar17} datapoints, it can be seen that the
leading stream in the model and data begin to diverge past the leading apocenter
as the stream returns to the Galactic plane.  The model stream pierces the plane at about
$X \approx 8\kpc$ in heliocentric coordinates ($X \approx 16 \kpc$ in galactocentric coordinates),
versus a modestly extrapolated position
of $X \approx 15 \kpc$ for the observed stream (see also \citealp{newberg07}).
In tandem, the velocity trend begins to
diverge from the trend found by \citetalias{law10} or \citet{belokurov14} at a similar longitude.
This is one area where the \citetalias{law10} or \citet{law05} prolate model performs
somewhat better than Model~A.  In part this is because we did not explicitly fit this region.
However, limited experiments showed us that a perfect fit to the extreme leading stream
is not simply a matter of including more constraints.  The younger dynamical state of the
debris in our model appears to make it more difficult to fit this region, due to the larger
difference between the stream and the orbit in our model.

\begin{figure}
\onecol{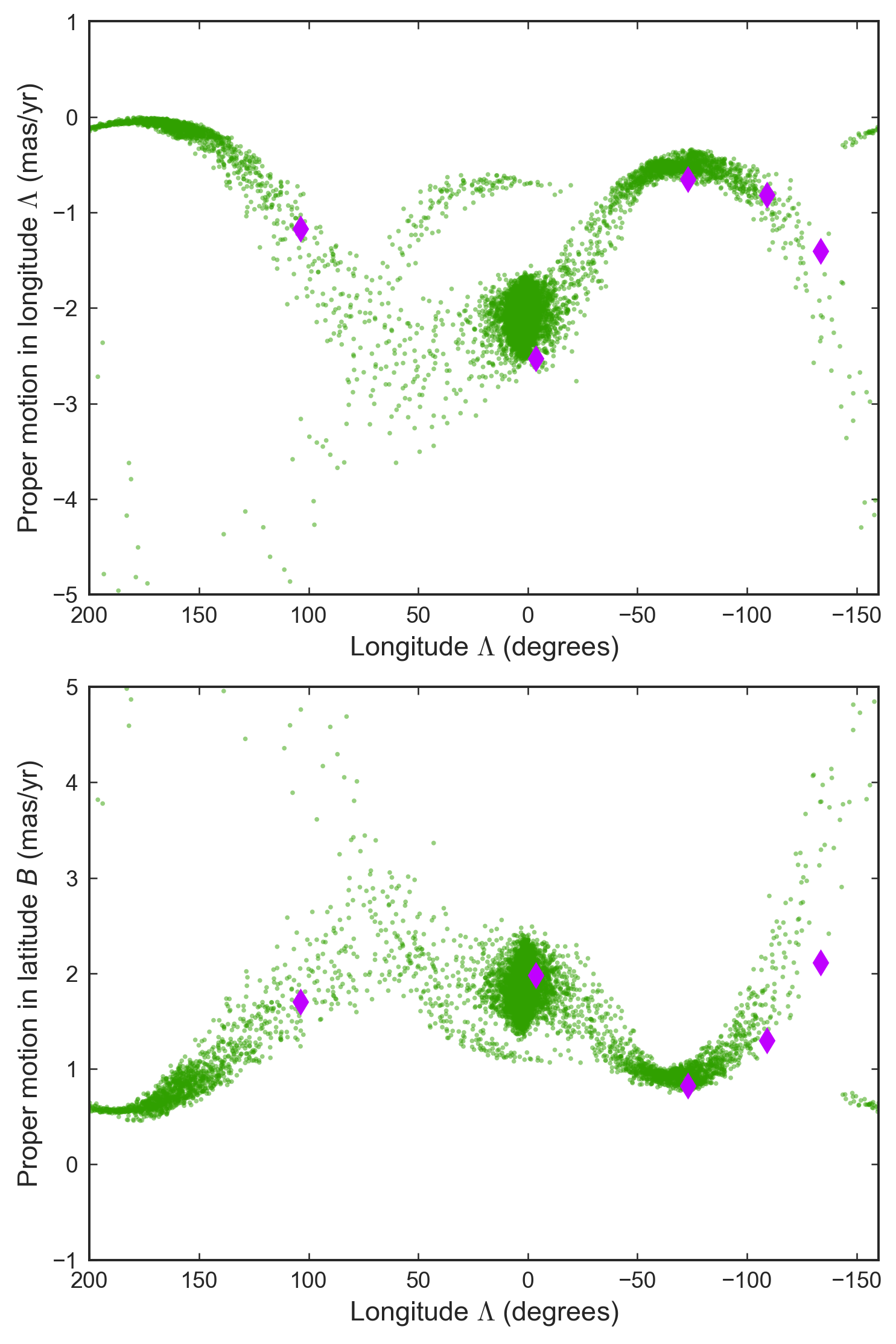}
\caption{
\label{fig.pm}
Proper motion of the stream along the longitude and latitude ($\Lambda$ and $B$)
directions, defined according to the convention of \citet{majewski03} and \citetalias{law10}.
The purple points show the values from \citet{sohn15,sohn16} converted to these axes.
}
\end{figure}

Fig.~\ref{fig.pm} shows the proper motion measured along the stream longitude
and latitude for the same \nbody\ model.  The use of these axes somewhat simplifies
the expected pattern compared to that seen using ecliptic or Galactic coordinates.
The solar motion adds a large component to both axes---if the Sun were at rest,
the proper motion in latitude would be insignificant.  We include the
precise proper motion results of \citet{sohn15,sohn16}.
These fall at least within the range of particle proper motions at each
location, although some points are noticeably offset from the mean values.
Note we did not use proper motion as a constraint in fitting the stream.
In examining this plot across our ensemble of models
we find in general that the success
in reproducing the leading stream's proper motion 
corresponds closely to the success in 
reproducing the leading stream distances and velocities.
Spherical potential models
generally perform badly with all three measures (in keeping with earlier work
such as \citealp{johnston05}),
as the leading stream returns too close or even interior
to the Sun in these cases, but flattening along the $y$ and $z$ axes can
produce reasonable agreement with observations.

We note that older and younger components are often offset
in the proper motion diagram, even though they do not generally separate cleanly.
This may explain some of the proper motion substructure found by \citet{sohn15}.
A full comparison, however, would require directly comparable analysis methods for
observations and simulation, and is outside the scope of this work. 

\begin{figure}
\onecol{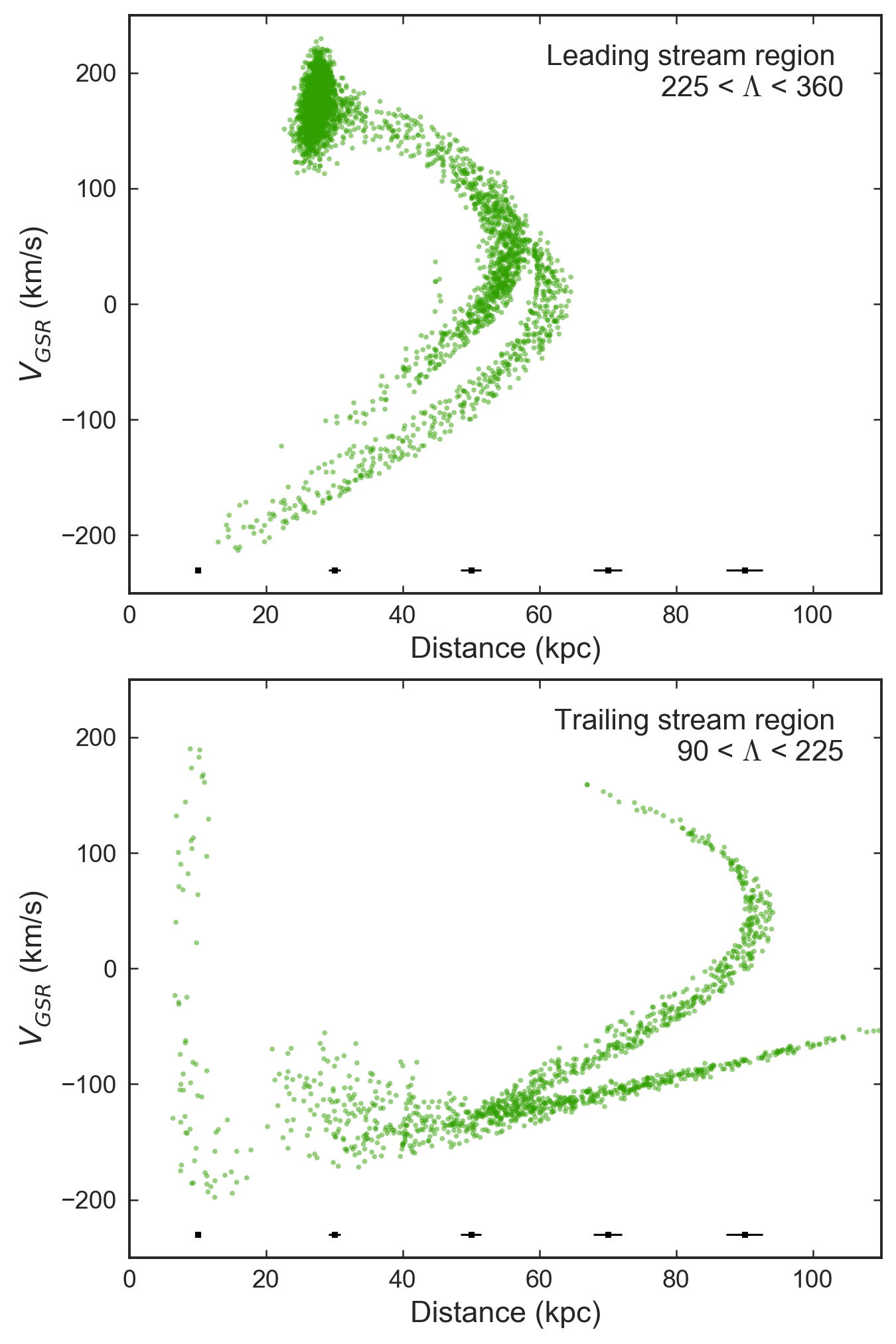}
\caption{
\label{fig.distvel}
Galactic standard of rest line-of-sight velocity from model A,
versus line-of-sight distance to the stream stars. 
We show two regions dominated by the leading and trailing stream respectively.
In these plots, the old and young stream components split into well-separated tracks
along much of the stream's range.  The separation appears cleaner than
in plots of either quantity versus longitude, as in Figs.~\ref{fig.distance} 
or \ref{fig.vel}.
Black points with error bars indicate the estimated distance error
of 3\% in the RR Lyrae sample of \citetalias{sesar17}, showing that the separation is
easily resolvable in principle.  
}
\end{figure}

Fig.~\ref{fig.distvel} shows the distance versus velocity for the stream particles,
in the leading and trailing regions separately.
Interestingly, the young and old components appear to separate better in this plot
than when either quantity is plotted against stream longitude, as in
Figs.~\ref{fig.distance} or \ref{fig.vel}.  
This is true in every case we have examined, regardless of whether we
use spray models or \nbody\ models, spherical or non-spherical models,
or runs with or without dynamical friction.
Apparently the azimuth is smeared out, likely by variations in the angular
momentum of the stars, in a way that does not affect the tight regularities
in the radial motion.
The separation between components is much clearer here
than in the model of \citetalias{law10},
because the debris in that model is dynamically older
and consists of more overlapping stream components.
Thus an observational version of this diagram would be a
powerful diagnostic of the stream's dynamical state.  
We find that proper motion also separates well in
some regions when plotted versus distance, although the regions with clear
offsets are more limited than in Fig.~\ref{fig.distvel}.

The motion in the young trailing tail of the stream probes the Milky Way halo
to the largest radius possible.  The observations of this young tail
already extend to Galactocentric radius $\approx 130 \kpc$.
According to \citetalias{sesar17}, the sample completeness is falling strongly
at this distance, so the actual tail probably extends even further.
Furthermore, in our models
the stars at these radii are typically infalling
from their individual apocenters several tens of kpc further out.
Thus the young trailing debris may well
probe the halo potential at even larger distances, up to $\tsim 150 \kpc$.
Obtaining reliable distances and velocities of stellar tracers to enable comparison to
Fig.~\ref{fig.distvel} would be extremely interesting.
We plan to address the information content of the young trailing stream in
future work.

As viewed on the sky, the observed stream has a ``bifurcation'' in the leading arm,
containing a denser branch that deviates from latitude zero towards the south (positive $B$)
and a fainter and narrower branch that remains near zero latitude \citet{belokurov06}.
The fainter branch is reported to lie closer by anywhere from 1 to 15 kpc
\citep{belokurov06,ruhland11,hernitschek17}.
A similar bifurcation has been reported in the trailing tail in the southern
hemisphere, though the structure is less clear \citep{koposov12,slater13}.
The latitude distribution of the model stream near leading
and trailing apocenter is centered on latitude $B \approx 0$, and has
an overall dispersion of about $6\deg$.  No clear bifurcation is apparent, but 
the model does supply at least some of the ingredients for explaining such
a bifurcation.  The leading stream consists of two distinct physical components,
and the one that is fainter past leading apocenter, namely the young stream,
also lies closer (see Fig.~\ref{fig.sprayex}).  For the southern hemisphere
observations of \citet{slater13} with $\Lambda \approx 110\degree$,
the older and closer one would likely be fainter, again consistent with observations.
Stellar population differences between the southern branches
noted by \citep{koposov12} may also be consistent with this picture.
What is currently missing from the model is a means to kick these two components
onto different orbital planes.
This could include a more complicated potential shape that affects the paths
traversed by young and old components in a differential fashion;
an encounter with a perturbing satellite;
or coherent stellar motions in the progenitor \citep{jorge10,gibbons16}.
The latter explanation is perhaps the most attractive in the context of
our model, since it could also produce narrow latitude distributions of
the individual components resembling those observed, in contrast to
the single broad distribution resulting from our hot spherical progenitor.

By measuring the dispersion in radial velocity around the overall trend
in the southern trailing tail ($25\degree < \Lambda < 90\degree$),
we obtain a velocity dispersion of $12 \kms$ in Model A.
Given that \citet{gibbons17} found dispersions of 
$8$ and $13 \kms$ in the metal-rich and metal-poor components
respectively, this seems reasonable.  We note that we chose the initial satellite
mass largely to produce a reasonable
velocity dispersion, so this value is in no way a surprise.

Although we are not advocating a single best fit to the Galactic potential here, some comment
on the potential is worthwhile.  In Model A, the virial radius
(defined here as the radius where the mean enclosed density is 100 times the critical density)
is $r_{100} = 331 \kpc$, and the virial mass is $\log M_{100} = 12.29$.
Over the {\it BDH-qyqz} sample, we find virial masses of $\log M_{100} = 12.34 \pm 0.06$.
For the {\it BDH-qyqz-DF} sample the results are similar at $\log M_{100} = 12.31 \pm 0.10$.
These estimates are on the high side compared to some estimates of the Milky Way mass,
but in agreement with or even low compared to some others
\citep{watkins10,gnedin10,vandermarel12,bland-hawthorn16}.
In particular, combining the Milky Way stellar mass from \citet{licquia15} with
the cosmological stellar mass-halo mass relation of \citet{behroozi10} would yield a
virial mass of $\log M_{100} = 12.5$.
However, our virial mass estimates, like many other methods, involve 
extrapolations based on an assumed form for the Milky Way potential,
and thus are highly uncertain.
The mass within 60~kpc over the {\it BDH-qyqz} sample is $(4.1 \pm 0.4) \times 10^{11} \msun$,
with similar results for most other successful runs.
This is in good agreement with the estimate of $(4.1 \pm 0.7)  \times 10^{11} \msun$
from \citet{xue08}.
The mass within 100~kpc over the {\it BDH-qyqz} sample s $(7.1 \pm 0.7) \times 10^{11} \msun$,
with similar results for most other runs.
This is markedly higher than the estimate 
$(4.0 \pm 0.7) \times 10^{11} \msun$ at the same radius, obtained
by \citet{gibbons14} through fitting the Sgr stream with a less flexible potential.

A more troubling issue is the scale radius of the NFW halo, which
in Model~A is 68~kpc.
This scale length is about twice the mean value
expected from the concentration-virial mass relation
and is near the upper extreme of the distribution \citep{neto07}.
We note that some alternative halo parameterizations, such as the Einasto form,
would yield a less strongly downcurving rotation curve,
and could move the turnover in the halo rotation curve to smaller radius.
Adjustments to our model such as stronger dynamical friction could also allow smaller scale lengths.
Effects of baryonic physics on Milky Way-like halos are still under study,
though whether the scale lengths can increase well beyond
those inferred from dark-matter-only simulations remains to be seen.
We discuss in the next section other effects,
such as the dynamical effect of the Large Magellanic Cloud (LMC),
that could affect the preferred potential.
\section{DISCUSSION}
\label{sec.discussion}
\subsection{Comparison to other models}
Many models of Sagittarius have appeared in the literature,
including those of
\citet{law05,fellhauer06,jorge10,purcell11,gibbons14} and
\citet{gomez15},
and it is beyond the scope of this paper to discuss them all.
However, two in particular have illustrated good agreement with several
stream observables as well as relevance to the issues discussed in this
paper, namely \citetalias{law10} and \citetalias{dierickx17}.
\citetalias{law10} was an \nbody\ simulation conducted in a fixed potential,
generated through a careful process of fitting to the
stream observations available at the time.
\citetalias{dierickx17} was instead a simulation generated with a live Milky Way
halo and thus included dynamical friction.  The potential was not fitted to
the stream, but specified a priori based on physical arguments, and the 
Sagittarius satellite properties were used
to constrain the orbit all the way back to first infall past the virial radius.
(We will not discuss here the followup paper \citealp{dierickx17b}, which
somewhat extends the study of these past orbits
but contains no new detail about the structure of the stream itself.)

The \citetalias{law10} model for the stream reproduces the
leading and southern trailing arms very well, but completely fails in the vicinity of
the trailing apocenter (the extent of which was not clear at the time).
In this model the trailing and leading apocenters are at
galactocentric radii of roughly 67 and 48 kpc (vs 100 and 50 kpc observed).
In contrast, the \citetalias{dierickx17} model is a worse match to many of the stream
observables; besides the typical failure to reproduce the leading stream,
it seems to be tilted by roughly $30 \degree$ from the stream plane.
It does come far closer to reproducing the trailing apocentric
distance than \citetalias{law10}, though the apocenters are
both slightly too high, indicative of an excessive orbital energy.
It also reproduces the outer trailing tail and leading ``fluff''
(features 2 and 3 from \citetalias{sesar17}) quite clearly,
showing a closer match to the observed structure in this respect
than the \citetalias{law10} model.

Although \citetalias{law10} contains a bulge and disk and its halo is triaxial,
the halo radial profile is logarithmic and in consequence the rotation curve is
basically flat.  \citetalias{dierickx17} instead uses a Hernquist form for the halo,
and as a result the rotation curve is falling all the way from the inner few kpc.
This raises an obvious question.
We have said that the more steeply the rotation curve falls (at least in the range
between the two apocenters), the smaller the ratio of trailing to leading apocentric
distances will be.  Yet the \citetalias{law10} and \citetalias{dierickx17} models seem to exhibit
the opposite trend.  What is the explanation?

The key point is that our argument is valid for streams
originating from specified pericentric passages.  In contrast,
the debris in the \citetalias{law10} and \citetalias{dierickx17}
models come from different pericentric passages.
In \citetalias{dierickx17} we can just count the trailing streams
pericentric passages in their fig.~8; the rounded stream at
the first trailing apocenter is from the second-to-last full passage,
just as in our models.  The discrete streams are less obvious in \citetalias{law10},
but we have resimulated their model with our spray code and found
that most debris at trailing apocenter is from the third-to-last full
passage, i.e. one to two orbits before that in \citetalias{dierickx17}.
The curved stubby tail at trailing apocenter
is from the second-to-last, not the last full passage, which explains
why it is so much more rounded than the young straight trailing tail
in \citetalias{dierickx17} or in our runs.

One reason \citetalias{law10} involves older debris is that steeper
rotation curve falloffs tend to make the stream much more stretched,
while shallower falloffs as in \citetalias{law10} tend to compress the stream
along its length \citep{dubinski99}.
In addition, the larger Sagittarius mass in \citetalias{dierickx17} also makes the streams
from given pericenters extend more than in \citetalias{law10}.
These two factors cooperate
to make the \citetalias{law10} debris dynamically older, and thus make it follow
the progenitor orbit much more faithfully than in \citetalias{dierickx17} or our models.

The remaining factor that helps the \citetalias{dierickx17} model achieve a larger
trailing apocentric distance is dynamical friction, as it
was run in a live potential.
Their fig.~4 shows the time between pericenters dropped from
about 1.5 to 1.3 Gyr in the last two cycles, which by
Equation~\ref{eqn.vctratio_obs} should
allow the
ratio of circular velocities to be $\tsim 7$\% lower than it
would otherwise.  
The \citetalias{law10} model in contrast lacks dynamical friction as it was run
in a fixed potential.
We have estimated circular velocities and orbital timescales
for both these models, and they both seem in accord with
the timescale arguments of Sections~\ref{sec.spherical}--\ref{sec.dynfric}.

Although the \citetalias{dierickx17} model constructs the stars at leading and trailing
apocenter in a similar manner to ours, it includes four full pericentric passages, not just two.
The two oldest streams are fairly tenuous, perhaps because the mass loss in the
central baryonic component has not started in earnest,
but these faint streams continue well beyond the leading and trailing apocenters.
It is plausible that such streams are emitted during the early orbital evolution
of the Sagittarius dSph, before the period we have considered.
Due to its 8 Gyr timescale and the small radial 
period of the orbit, the \citetalias{law10} model stream also wraps far beyond the
limits to which the Sagittarius stream has currently been detected.  Some of
our models have highly extended streams even from our limited set of
pericentric passages, while others do not.  Clearly, detection of such
highly wrapped debris would make for a powerful constraint on Milky
Way models.

The stream modeling of \citet{gibbons14}, while an important contribution, does not provide 
the detailed plots or full specification that would allow a detailed comparison with
observations or with our own models.  The modeling techniques and observational
inputs used there are not very different than in this paper.  It therefore may seem
puzzling they derive a low mass for the Galactic halo, $5.6 \pm 1.2 \times 10^{11} \msun$
at $200 \kpc$, in contrast to our much higher preferred masses.  We have not found any
hint of contradiction between our modeling techniques.  However, the {\it TF}
model used in that paper requires a steep falloff in the rotation curve in order to
fit the azimuth of the leading and trailing apocenters.  Our best-fitting models
have the flexibility to bend the rotation curve back upwards to fit the apocenter
ratio better, but their 3-parameter potential does not, and therefore their mass
at large radius is very low.  The lesson we take
from this is not that one model is more correct than another.  Rather, we infer
that seemingly small issues within the likelihood function and the degrees of freedom in
a tidal stream model can drive large differences in the model implications.  This is
especially true when definite discrepancies remain between models and observations
(cf.\ the prolate-oblate debate spurred by the Sgr leading stream).
As the observational dataset improves, we believe our understanding will be improved best by
looking for features that will help eliminate degrees of freedom from the modeling.
The distinct components visible in Figure~\ref{fig.distvel} are examples of such features.

\subsection{Future directions}
Overall, it seems fair to say Model A is the closest that any published model 
has come to reproducing the various observables of the Sagittarius stream.
The model does almost as well as \citetalias{law10} in reproducing the
leading stream, and much better with the more distant parts of the trailing stream.
Furthermore, it agrees significantly better with the observed
orbital plane orientation, distance values, and distance and velocity dispersions
than the model of \citetalias{dierickx17}.
However, we deliberately refrain from calling this the
``best fit'' to the data in any quantitative sense, for three reasons.
The first is that there are some discrepancies with observations still:
most prominently the velocity in the trailing stream, the shape and velocity of
the leading arm as it returns to the Galactic plane, and 
the complex, bifurcated latitude distribution of the observed stream.
The definition of a best fit in this case is highly dependent on the choice of
input data and the exact construction of a likelihood function.

The second reason is that we know there are many things missing from our comparison
with observations.
We have taken the stream's latitude and proper motion into account only 
in a qualitative way (to argue in favor of non-spherical models like Model~A),
and we have not addressed the bifurcation in latitude.
Neither have we used the motion of Sagittarius itself in fitting the stream.
We used arbitrarily inflated errors on individual points rather than 
carefully taking into account possible systematic errors in
the observations, e.g. in the assigned RR Lyrae distance scale
of \citetalias{sesar17}.  
We have not approached the analysis of the observed stars and the simulations
in an equivalent fashion.
Finally, we expect new data to arrive soon from the {\it Gaia} survey
which will greatly augment the current dataset and quickly render obsolete
any current judgement about the best fit to the data.

A third reason is that we know there are other physical effects that
we have ignored.  Primary among these is the tidal force from the
Large and Small Magellanic Cloud system, which most likely has
experienced its first close encounter with the Milky Way only $\ltrsim
200 \Myr$ ago \citep{nitya13}, and can have an effect on the
Sagittarius stream that varies from negligible to substantial
depending on its mass and orbit
(\citetalias{law10}; \citealp{vera-ciro13}; \citealp{gomez15}).
We have conducted preliminary experiments with an LMC tidal force,
and so far found that it does not greatly affect the apocenter radii of the
stream, but more work is required to understand its full effects on the models.
It is also unreasonable to expect the dark halo's potential to have a constant
ellipsoidal distribution with radius, or be arbitrarily aligned along
the $X$/$Y$/$Z$ directions, as adopted in our simple models.
There are many other smaller effects we have neglected, including possible
perturbations by the Galactic bar, perturbations from the Milky Way's
other known satellite galaxies or unknown dark subhalos,
and growth of the Milky Way's potential with time.
Finally, and perhaps most importantly,
our model of Sagittarius is a single, hot,
spherical component, with no distinction between stellar and dark distributions.
A dark halo can amplify the effect of dynamical friction even after
several epochs of tidal stripping, while
cold coherent stellar motions in the progenitor
can result in significant effects on the
position and velocity dispersion of the stream \citep{jorge10}.
We aim to remedy some of these deficiencies in future work.

We expect future progress in understanding the Sagittarius stream
to come from new observations, not just by providing more precise error bars
for various quantities, but by also making clear the nature of morphological
features seen in the stream.  For example, we may be able to distinguish
debris from different pericentric passages in velocity, latitude, and
proper motion spaces, as well as distance, which would help eliminate
several dimensions of freedom from the current range of models.
A crucial issue is whether the outer ``fluff'' in the leading stream
and/or the ``outer Virgo overdensity'' are
actually identifiable as an older component of Sagittarius.
\section{Conclusions}
\label{sec.conclusions}
In this paper, we have presented models approximating the behavior of
the Sagittarius stream, with a focus on reproducing the very different
apocentric distances of the leading and trailing arms and the substructure
within the stream.  We have found reasonable fits in models where the
satellite has just experienced its third disruptive pericentric passage,
and debris from the first and second passages make up the bulk of the stream.
We found streams from earlier pericentric passages to form the stars at apocenter
to be impractical, as then the ratio of apocenters is too small.
Within the class of models considered, we
have found simple relationships connecting the orbital timescales at the
apocenters to the apocentric distances and circular velocities.
These can be used to constrain the shape of the rotation curve and the
influence of dynamical friction.   

Perhaps unexpectedly,
the agreement with models is best when the rotation curve has
an upward curvature: our most successful models fall from 10 to 50 kpc but are
roughly flat from 50 to 100 kpc.  We caution that this result is
tentative, and can be modified by other factors.
Dynamical friction affects the required rotation curve by changing
the ratio of timescales experienced by the stars at leading and trailing apocenter.
If we can contrive to make dynamical friction stronger
without boosting the velocity dispersion in the stream too much,
it would allow a steeper falloff in the rotation curve between 50 and 100 kpc.
Fortunately, it should be possible to break this degeneracy between
potential shape and dynamical friction by disentangling
the young and old streams at the leading apocenter.
Potentials with more complex shapes may alter the best shape
of the rotation curve.
Prolate halos are helpful for matching the leading stream's properties,
but they increase the pressure for an upward-bending galactic potential.
We also caution that we have neglected here
several important physical effects, such as the influence of the LMC.

Our focus has been on physical regularities and morphological features
that can be used to interpret future observations of the Sgr stream,
rather than statistical description of fits to the current data.
Overall, however, our results point to a more extended and massive galactic 
halo than used in standard Galactic models.
This conclusion can be tested in the distinct outer trailing tail of
the stream, which probes distances well beyond 100 kpc 
where existing dynamical tracers are extremely sparse.
Future observations of the Sagittarius stream will thus be able to
measure the mass of our galaxy to unprecented distances.
\begin{table}
\caption{Distances derived from Sesar et al.\ (2017) RR Lyrae sample.
$\Lambda$ gives the central longitude of the equal-width bins.
``Statistical error'' refers to the formal uncertainty obtained while fitting the binned points.
``Adopted MCMC error'' is the increased uncertainty we adopted in our likelihood function
to ensure a wide exploration of parameter space in the MCMC runs.}
\label{tab.distance}
\begin{tabular}{llll}
\hline
$\Lambda$ & Distance & Stat.\ error & Adopted MCMC error\\
($\degree$) & (kpc) & (kpc) & (kpc)\\
\hline
\multicolumn{4}{c}{Leading}\\
\hline
266.0 & 39.4 & 4.0 & 5.9 \\
274.0 & 46.5 & 1.3 & 2.6 \\
282.0 & 48.5 & 0.9 & 2.4 \\
290.0 & 50.9 & 0.8 & 2.5 \\
298.0 & 52.1 & 0.7 & 2.4 \\
306.0 & 50.4 & 0.8 & 2.4 \\
314.0 & 47.0 & 1.0 & 2.5 \\
\hline
\multicolumn{4}{c}{Southern trailing}\\
\hline
123.0 & 31.6 & 1.5 & 2.6 \\
131.0 & 35.8 & 2.3 & 3.6 \\
139.0 & 38.1 & 2.2 & 3.5 \\
147.0 & 47.9 & 2.5 & 4.0 \\
\hline
\multicolumn{4}{c}{Northern trailing}\\
\hline
179.0 & 91.0 & 2.0 & 4.8 \\
187.0 & 93.1 & 2.1 & 4.9 \\
195.0 & 89.9 & 1.4 & 4.3 \\
203.0 & 85.1 & 1.7 & 4.4 \\
211.0 & 80.9 & 1.9 & 4.4 \\
219.0 & 74.9 & 3.4 & 5.8 \\
\hline
\end{tabular}
\end{table}

\begin{table*}
\caption{Velocity measurements in used in fitting, selected from \citet{belokurov14}.
As with the distance data, we increased the uncertainty over the formal statistical
uncertainty for use in our MCMC runs.}
\label{tab.velocity}
\begin{tabular}{llll}
\hline
$\Lambda$ ($\degree$)&GSR velocity ($\mbox{km}\,\mbox{s}^{-1}$)&Stat.\ error&Adopted MCMC error\\
\hline
\multicolumn{4}{c}{Leading}\\
\hline
260.4 & -70.7 & 4.4 & 20.4 \\
266.5 & -60.2 & 3.1 & 19.2 \\
272.7 & -29.5 & 6.0 & 22.3 \\
278.8 & -16.7 & 6.7 & 23.3 \\
285.0 & -7.6 & 5.4 & 21.6 \\
291.2 & 8.8 & 3.2 & 19.3 \\
297.3 & 41.1 & 9.2 & 27.2 \\
\hline
\multicolumn{4}{c}{Southern trailing}\\
\hline
67.5 & -7.8 & 2.8 & 19.0 \\
75.0 & -35.4 & 1.6 & 18.2 \\
82.5 & -58.8 & 2.1 & 18.5 \\
87.5 & -71.8 & 1.3 & 18.1 \\
92.5 & -87.2 & 1.2 & 18.1 \\
97.5 & -98.6 & 1.2 & 18.1 \\
102.5 & -108.8 & 1.4 & 18.2 \\
107.5 & -120.0 & 1.2 & 18.1 \\
112.5 & -129.5 & 2.1 & 18.5 \\
117.5 & -135.1 & 1.6 & 18.2 \\
122.5 & -141.9 & 1.8 & 18.3 \\
127.5 & -150.8 & 3.6 & 19.6 \\
132.5 & -141.1 & 2.7 & 18.9 \\
142.5 & -127.2 & 3.1 & 19.2 \\
\hline
\multicolumn{4}{c}{Northern trailing}\\
\hline
180.6 & -31.3 & 16.8 & 41.6 \\
187.6 & -13.8 & 16.8 & 41.6 \\
194.7 & 17.8 & 5.5 & 21.7 \\
201.7 & 44.7 & 2.0 & 18.4 \\
208.8 & 77.0 & 3.6 & 19.6 \\
215.8 & 128.8 & 2.3 & 18.6 \\
222.8 & 132.5 & 3.6 & 19.6 \\
\hline
\end{tabular}
\end{table*}

\begin{table*}
\caption{Models tested with MCMC runs.  Median likelihood obtained from
the final state sample are included (smaller numbers are better).}
\label{tab.models}
\begin{tabular}{lllll}
\hline
Model class & Median likelihood & $\log_{10} M_{sat}$ \\
\hline
\multicolumn{5}{c}{Near-spherical models, no dynamical friction}\\
\hline
{\it galpy2014} & -29.6 & 9.1\\
{\it TF} & -18.5 & 9.1\\
{\it 1PL} & -10.5 & 9.1\\
{\it 2PL} & -7.6 & 9.1\\
{\it BDH-sph} & -6.3 & 9.1\\
\hline
\multicolumn{5}{c}{Near-spherical models, with dynamical friction}\\
\hline
{\it galpy2014-DF} & -27.1 & 9.1\\
{\it TF-DF} & -11.0 & 8.5\\
{\it 1PL-DF} & -7.3 & 9.1\\
{\it 2PL-DF} & -6.8 & 9.1\\
{\it BDH-sph-DF} & -6.2 & 9.1\\
\hline
\multicolumn{5}{c}{Aspherical models}\\
\hline
{\it BDH} & -7.0 & 9.1\\
{\it BDH-qz} & -5.4 & 9.1\\
{\it BDH-qyqz} & -5.4 & 9.1\\
{\it BDH-qyqz-DF} & -5.9 & 9.1\\
\hline
\end{tabular}
\end{table*}

\begin{table*}
\setlength{\tabcolsep}{3pt}
\caption{Selected model states from variants of the bulge-disk-halo models.
$v_{tan,gsr}$ is the current tangential velocity and $d_{Sgr}$ the heliocentric
distance of the Sgr dwarf.  This results in the current dSph position and velocity
given by $x_{gc}$, $y_{gc}$, $z_{gc}$, $v_{x,gsr}$, $v_{y,gsr}$, and $v_{z,gsr}$.
The simulation starts the dSph along its orbit at lookback time $t_{ev}$.  
The initial outer radius of the King model profile is given by $r_{outer}$.  
As detailed in Section~\ref{sec.potential}, 
$f_d$, $f_L$, and $f_M$ are the free parameters describing the mass and radial behavior.
$q_z$ is the flattening parameter along the $z$ axis.
The corresponding $y$-axis parameter is held fixed at 
$q_y= 1.1$ for the {\it BDH-qyqz} potential and $1$ otherwise.
Lengths are in kpc, velocities in km s$^{-1}$, and evolution time in Myr.}
\label{tab.states}
\begin{tabular}{llllllllllllllll}
\hline
Name &
Family&
$v_{tan,gsr}$&
$d_{Sgr}$&
$x_{gc}$&$y_{gc}$&$z_{gc}$&
$v_{x,gsr}$&$v_{y,gsr}$&$v_{z,gsr}$&
$t_{ev}$&
$r_{outer}$&
$\log f_d$&
$\log f_l$&
$\log f_M$&
$q_z$
\\
\hline
A&
{\it BDH-qyqz}&
229.65&
28.523&
19.235&
 2.683&
-6.981&
224.59&
-31.91&
174.58&
2938.2&
6.1642&
0.4857&
1.4431&
1.4862&
1.1061\\
B&
{\it BDH-qyqz-DF}&
234.60&
26.684&
17.460&
 2.510&
-6.530&
225.87&
-32.95&
179.26&
2957.6&
5.7552&
0.5530&
1.7448&
1.9495&
1.1181\\
C&
{\it BDH-qz}&
217.67&
26.960&
17.727&
 2.536&
-6.598&
221.48&
-29.41&
163.30&
2771.9&
6.3479&
0.3358&
1.8166&
1.9772&
1.1223\\
D&
{\it BDH}&
248.75&
25.997&
16.798&
 2.446&
-6.362&
229.54&
-35.91&
192.59&
2629.4&
6.0433&
0.5339&
1.5155&
1.5687&
1\\
E&
{\it BDH-sph}&
245.14&
24.919&
15.757&
 2.344&
-6.098&
228.61&
-35.15&
189.19&
2325.3&
6.0486&
0.4247&
1.6436&
1.7768&
1\\
\end{tabular}
\end{table*}

\section*{ACKNOWLEDGMENTS}
We thank Tom Quinn and Joachim Stadel for the use of PKDGRAV, 
and Josh Barnes for the ZENO package used in building the initial conditions.
We thank Jo Bovy for making public the {\tt galpy} package and assisting with its use,
Martin Weinberg assisted with use of the \textsc{BIE} code and provided
a routine to initialize King models.
This research made use of the open-source Python packages {\tt astropy} \citep{astropy18},
{\tt numpy}, and {\tt matplotlib}.
We thank Martin Weinberg, Gretchen Zwart, and Eric Winter for assistance
with computing resources.
Support for this work was provided by NASA through grants for programs
GO-12564,
GO-13443, 
GO-14235,
and AR-15017
from the Space Telescope Science Institute (STScI), which is
operated by the Association of Universities for Research in Astronomy
(AURA), Inc., under NASA contract NAS5-26555.
HWR acknowledges support through the German Science Foundation 
(DFG) grant ``SFB 881: The Milky Way System (A3)''.
\footnotesize{
\bibliographystyle{mnras}
\bibliography{ms}
\label{lastpage}
}

\end{document}